\documentclass[journal, twoside, final]{IEEEtran}

\usepackage{amsmath,amssymb,amsfonts}
\usepackage{algorithmic}
\usepackage[ruled]{algorithm2e}
\usepackage{array}
\usepackage[caption=false,font=normalsize,labelfont=sf,textfont=sf]{subfig}
\usepackage{textcomp}
\usepackage{stfloats}
\usepackage{url}
\usepackage{verbatim}
\usepackage{graphicx}
\usepackage{cite}
\usepackage{booktabs}
\usepackage{makecell}
\usepackage{multirow}

\newtheorem{lemma}{Lemma}
\newtheorem{remark}{Remark}

\begin{document}

\title{Secrecy Sum-Rate Maximization for Active IRS-Assisted MIMO-OFDM SWIPT System}

\author{Xingxiang Peng, Peiran Wu, \IEEEmembership{Member,~IEEE}, Junhui Zhao, \IEEEmembership{Senior~Member,~IEEE}, and \\ Minghua Xia, \IEEEmembership{Senior~Member,~IEEE}

\thanks{Received 26 March 2024, revised 6 July 2024, accepted 6 October 2024. This work was supported in part by the National Natural Science Foundation of China under Grant 62171486 and Grant U2001213 and in part by the Guangdong Basic and Applied Basic Research Project under Grant 2022A1515140166. The associate editor coordinating the review of this paper and approving it for publication was K. Shen. \textit{(Corresponding authors: Minghua Xia; Peiran Wu.)}}

\thanks{Xingxiang Peng, Peiran Wu, and Minghua Xia are with the School of Electronics and Information Technology, Sun Yat-sen University, Guangzhou 510006, China (e-mail: pengxx8@mail2.sysu.edu.cn; wupr3@mail.sysu.edu.cn; xiamingh@mail.sysu.edu.cn).}

\thanks{Junhui Zhao is with the School of Electronic and Information Engineering, Beijing Jiaotong University, Beijing 100044, China (email: junhuizhao@bjtu.edu.cn).}

\thanks{Color versions of one or more figures in this article are available at https://doi.org/10.1109/TWC.2024.3478252.}
	
\thanks{Digital Object Identifier 10.1109/TWC.2024.3478252}
}

\maketitle


\begin{abstract}
The propagation loss of RF signals is a significant issue in simultaneous wireless information and power transfer (SWIPT) systems. Additionally, ensuring information security is crucial due to the broadcasting nature of wireless channels. To address these challenges, we exploit the potential of active intelligent reflecting surface (IRS) in a multiple-input and multiple-output (MIMO) orthogonal frequency division multiplexing (OFDM) SWIPT system. The active IRS provides better beamforming gain than the passive IRS, reducing the ``double-fading" effect. Moreover, the noise introduced at the active IRS can be used as artificial noise (AN) to jam eavesdroppers. This paper formulates a secrecy sum-rate maximization problem related to precoding matrices, power splitting (PS) ratios, and the IRS matrix. Since the problem is highly non-convex, we propose a block coordinate descent (BCD)-based algorithm to find a sub-optimal solution. Moreover, we develop a heuristic algorithm based on the zero-forcing precoding scheme to reduce computational complexity. Simulation results show that the active IRS achieves a higher secrecy sum rate than the passive and non-IRS systems, especially when the transmit power is low or the direct link is blocked. Moreover, increasing the power budget at the active IRS can significantly improve the secrecy sum rate.
\end{abstract}


\IEEEpubidadjcol

\section{Introduction} 
\label{Sec-Intro}
\IEEEPARstart{T}{he} continually increasing number and diversification of intelligent devices in Internet of Things (IoT) networks enhance the quality of human life by facilitating various applications such as smart cities, autonomous driving, and multisensory virtual reality \cite{9799791}. One of the main challenges in deploying reliable IoT is the energy limitation of battery-powered devices. To address this limitation, simultaneous wireless information and power transfer (SWIPT), leveraging the dual functionality of radio-frequency (RF) signals as carriers of both information and energy, has recently sparked a surge of research interest \cite{6489506}. Current investigations on SWIPT can be categorized into three main groups, including SWIPT systems with separated energy-harvesting (EH) and information-decoding (ID) receivers, e.g., \cite{6860253}; SWIPT systems with time-switching (TS)-based receivers, e.g., \cite{8636993}; SWIPT systems with power-splitting (PS)-based receivers e.g., \cite{9847302}. Besides, the rate-energy regions of these three types of SWIPT have been investigated in \cite{7934322}, and it was recognized that the PS-based SWIPT achieves a better tradeoff between the rate and the harvested energy than the TS-based one.

Besides energy shortage, information security is another pivotal issue in IoT networks due to the broadcasting nature of wireless channels. Note that, compared with traditional wireless networks without EH, EH-enabled networks are more vulnerable to attacks from unexpected eavesdroppers. This occurs because, due to the rapid attenuation of RF signals, the transmission power needs to be sufficiently high to meet the EH requirements of legitimate users, which concurrently raises the risk of information leakage to eavesdroppers \cite{7070727,7572901}. In this respect, physical layer security (PLS) techniques have been employed to guarantee information security in PS-based SWIPT systems \cite{7477999,7275174,7911267,7962221,8017508,8943160}. Specifically, the work \cite{7477999} investigated the robust beamforming and PS design for a multiple-input and single-output (MISO) SWIPT system, aiming to maximize the secrecy sum rate under both transmit and harvested power constraints. The work \cite{7275174} considered an orthogonal frequency-division multiple access (OFDMA) SWIPT network, in which the subcarrier allocation and the PS ratios were jointly designed to maximize the harvested energy while satisfying each user's individual secrecy rate constraint. The work \cite{7911267} studied the secrecy energy-efficiency maximization problem for a multiple-input and multiple-output (MIMO) SWIPT system and proposed an alternating optimization (AO)-based method to solve the formulated non-convex problem. In \cite{7962221}, the authors studied the transmit power minimization problem for a multi-user system under the target secrecy rate and harvested energy constraints. The work \cite{8017508} focused on the robust, secure transceiver design for MISO PS-based SWIPT systems under interference channels. Besides, the security issue for a MIMO orthogonal frequency-division multiplexing (OFDM) PS-based SWIPT system was investigated in \cite{8943160}.

\IEEEpubidadjcol

Intelligent reflecting surface (IRS) has recently emerged as a promising technology for future 6G wireless communications \cite{9446526}. IRS is a planar array composed of many low-cost, passive, and reconfigurable units, where each unit can reflect the incident signal with an adjustable phase shift \cite{10065384}. By carefully adjusting the phase shifts of the IRS, the received signal power of the legitimate users can be strengthened while that of the eavesdropper is suppressed. Thus, the secrecy performance is enhanced. Motivated by this, many works have been devoted to amalgamating PLS and IRS in conventional wireless communication systems, e.g., \cite{9446526,10065384,9903846,9520295}, and RF-EH-enabled wireless communication systems, e.g., \cite{9234391,9288742,9618858,9975289,9913501,10354056,9766101,9734045,10288203,10547445}. However, although IRS brings a new reflection link for signal transmission in addition to the direct link, a ``double-fading'' effect emerges in this reflection link, i.e., the signals received via this link suffer from large-scale fading twice \cite{9998527}. 

To mitigate the ``double-fading'' effect, a new IRS architecture known as the active IRS has recently been conceptualized. The active IRS and passive IRS mainly differ in three aspects \cite{9998527,9652031,9963962}: 1) Each reflecting element on an active IRS is additionally integrated with a reflection-type amplifier (e.g., a tunnel diode circuit); 2) Each reflecting element on an active IRS consumes additional power, and the noise introduced by the reflection-type amplifier cannot be neglected; 3) Each reflecting element on an active IRS is not only capable of adjusting the phase but also amplifying the amplitude of the incident signal. In conclusion, the active IRS requires higher hardware costs and additional power consumption than the passive IRS, but it alleviates the ``double-fading'' effect and offers greater configuration flexibility. Some innovative works have investigated the potential secrecy gain provided by the active IRS in wireless systems \cite{9652031,9963962,10061167,10275057}. Specifically, the works \cite{9652031,9963962,10061167} exploited the secure potential of active IRS in downlink MISO scenarios, while the work \cite{10275057} considered an uplink single-input and multi-output (SIMO) non-orthogonal multiple access (NOMA) network. However, these solutions are limited to pure communication systems and do not apply to a SWIPT system. Regarding most research contributions that exploited active IRS to SWIPT systems, e.g., \cite{9810984,9849458,10012424,10280714}, they did not consider the scenarios in the presence of eavesdroppers; thus, the security of the information transmitted over wireless networks cannot be guaranteed. To our best knowledge, the secrecy potential and the PLS design of integrating an active IRS into SWIPT systems still need to be investigated.

\begin{table*}[t] \tiny
	\caption{Comparison Between This Work and Related Studies}
	\label{table_cmp}
	\centering
	\begin{tabular}{|c|c|c|c|c|c|}
		\hline
		\textbf{Scenarios} & \textbf{Ref.} & \textbf{System Configurations} & \textbf{EH Protocol}  & \textbf{Objective Metrics} & \textbf{Solutions and Approaches}  \\
		\hline
		\multirow{10}{*}[-10ex]{ \makecell{\textbf{PLS design in} \\ \textit{\textbf{passive-IRS}} \\ \textbf{assisted systems}}} &\cite{9520295} &   MIMO, OFDM & w/o EH & secrecy sum-rate & \makecell{AO-based solution\\Lagrangian method, MM method, projection method} \\
		\cline{2-6} 
		& \cite{9234391} &  MIMO & SWIPT with separated EH and ID  & secrecy rate & \makecell{inexact BCD  (IBCD)-based solution\\MM method, complex circle manifold method} \\
		\cline{2-6}  
		& \cite{9288742} &  MISO & SWIPT with separated EH and ID  & harvested power & \makecell{AO-based solution, \textbf{low-complexity solution \checkmark}\\ SDR method, Gaussian randomization procedure}\\
		\cline{2-6} 
		& \cite{9618858} &  multi-user, MISO & SWIPT with separated EH and ID  & worst-case secrecy rate & \makecell{AO-based solution\\penalty dual decomposition, SCA method}\\
		\cline{2-6} 
		& \cite{9975289} &  MISO & SWIPT with separated EH and ID & worst-case secrecy rate & \makecell{BCD-based solution\\stochastic SCA, concave-convex procedure}\\
		\cline{2-6} 
		& \cite{9913501} &  multi-user, MIMO &  wireless-powered communication system & secrecy sum-rate & \makecell{AO-based solution, \textbf{low-complexity solution \checkmark}\\MM method, SCA method, penalty method}\\
		\cline{2-6} 
		& \cite{10354056} &  MISO & wireless-powered communication system & secrecy rate & \makecell{AO-based solution\\Charnes Cooper transform, SDR}\\
		\cline{2-6} 
		& \cite{9766101} &  MISO & PS-based SWIPT, nonlinear EH & secrecy rate & \makecell{AO-based solution\\Charnes Cooper transform, SDR}\\
		\cline{2-6} 
		& \cite{9734045} &  MISO &  PS-based SWIPT & secrecy rate & \makecell{AO-based solution, \textbf{deep learning-based solution \checkmark}\\feasible point pursuit SCA, penalty method}\\
		\cline{2-6} 
		& \cite{10288203} &  multi-user, MISO & PS-based SWIPT, nonlinear EH  & worst-case secrecy rate & \makecell{AO-based solution, \\SCA method, SDR method}\\
		\cline{2-6} 
		& \cite{10547445} &  multi-user, MISO, NOMA & PS-based SWIPT & secrecy sum-rate & \makecell{AO-based solution, \\ SCA method}\\
		\hline
		\multirow{5}{*}[-5ex]{ \makecell{\textbf{PLS design in} \\ \textit{\textbf{active-IRS}} \\ \textbf{assisted systems}}} & \cite{9652031} &  MISO &  w/o EH  & secrecy rate & \makecell{AO-based solution\\Charnes Cooper transform, SDR method, MM method}\\
		\cline{2-6} 
		& \cite{9963962} &  MISO & w/o EH & transmit power & \makecell{AO-based solution\\penalty method, SCA method}\\
		\cline{2-6} 
		& \cite{10061167} &  MU-MISO & w/o EH & secrecy sum-rate & \makecell{AO-based solution\\penalty dual decomposition, MM method}\\
		\cline{2-6} 
		& \cite{10275057} &  SIMO, NOMA & w/o EH & secrecy sum-rate & \makecell{BCD-based solution\\SCA method}\\
		\cline{2-6} 
		& This work &  MIMO, OFDM & PS-based SWIPT, nonlinear EH & secrecy sum-rate & \makecell{BCD-based solution, \textbf{low-complexity solution \checkmark} \\ SDR method, penalty method, IA method, ZF method}\\
		\hline
	\end{tabular}
\end{table*}

Against these backgrounds, this paper studies the PLS design for an active-IRS-assisted MIMO-OFDM SWIPT system. For clarity, the comparison between our work and the other publications on the PLS design of IRS-assisted wireless systems is shown in Table \ref{table_cmp}. In \cite{9520295}, it was assumed that only a single stream is transmitted on each subcarrier. Thus, the proposed solution cannot be applied to general spatial-multiplexing MIMO scenarios, as considered in this work. Compared to the works \cite{9234391,9288742,9618858,9975289,9913501,10354056,9766101,9734045,10288203,10547445}, we adopt the active IRS to achieve better secrecy performance and employ the OFDM modulation to combat the channel's frequency-selectivity. Compared to the works \cite{9520295,9652031,9963962,10061167,10275057}, the PS protocol is used in the system design, which enables the legitimate receiver to harvest RF energy from the received signal to prolong its lifetime. In a nutshell, the significant contributions of this work are summarized as follows:

\begin{enumerate}
	\item We maximize the secrecy sum rate of an active IRS-assisted MIMO-OFDM SWIPT system by jointly optimizing the information/artificial noise (AN) precoding matrices, PS ratios, as well as the IRS matrix. However, this problem is hard to solve since the matrix-based optimization variables are intricately coupled in the objective function and the constraints. To tackle this challenge, we transform the problem into a more tractable form and then employ the block coordinate descent (BCD) framework to solve the resulting problem. 
	
	\item  Within the BCD framework, we optimize the information/AN precoding matrices for given PS ratios and IRS matrix based on the semi-definite relaxation (SDR) method. For given information/AN precoding matrices and IRS matrix, the optimal solutions to the PS ratios are derived in semi-closed form with the aid of the penalty method. For given information/AN matrices and PS ratios, we develop an inner approximation (IA)-based algorithm to optimize the IRS matrix, which is guaranteed to converge to a Karush-Kuhn-Tucker (KKT) solution.
	
	\item To reduce the computational complexity of the BCD-based solution, we further design a heuristic algorithm based on a three-stage optimization strategy. First, we adjust the IRS matrix to maximize the channel gain of the legitimate user. Then, a zero-forcing (ZF)-based structure is established for the information precoding matrices, with which the original matrix-based optimization problem can be simplified into a scalar-based one that can be solved with low complexity. Finally, the PS ratios are optimized based on the IA method.
	
	\item Simulation results demonstrate that introducing an active IRS can significantly increase the secrecy sum rate, compared to passive- and non-IRS systems, especially when the transmit power is low or the direct link is blocked. Moreover, it is also shown that the low-complexity algorithm can achieve a performance comparable to that of the BCD-based algorithm.
\end{enumerate}

The rest of this paper is organized as follows: Section \ref{Sec-SysMod} introduces the system model and formulates the optimization problem. Sections \ref{Sec-BCD} and \ref{Sec-lowCompAl} propose BCD-based and low-complexity algorithms to solve the formulated problem. Section \ref{Simul} presents and discusses numerical results to evaluate the performance of the proposed algorithms. Finally, Section \ref{Sec-Concl} concludes the paper.

\textit{Notation:} In this paper, vectors and matrices are denoted by boldface lower- and upper-case letters, respectively. The symbol $\mathbb{C}^{M \times N}$ represents the $M \times N$ complex-valued domain. The operators $\vert \cdot \vert$ and $\|\cdot\|_{F}$ denote the Euclidean scalar norm and matrix Frobenius norm, respectively. The operators $(\cdot)^T$, $(\cdot)^H$, $(\cdot)^{-1}$, $\text{Tr}(\cdot)$ and $\text{det}(\cdot)$ denote the transpose, Hermitian transpose, inverse, trace, and determinant of a matrix, respectively. The $(m,n)$-th element of a matrix is denoted by $[\cdot]_{m,n}$. The operators $\odot$ and $\otimes$ represent Hadamard and Kronecker products, respectively. The operator $\mathbb{E}[\cdot]$ denotes the statistical expectation. The symbols $\text{diag}\{\cdot\}$, $\text{blkdiag}\{\cdot\}$, and $\text{blkcirc}\{\cdot\}$ denote the diagonal, block diagonal, and block circulant operations, respectively. The symbols $\Re\left\{\cdot\right\}$ and $\operatorname{arg}\{\cdot\}$ represent the real part and phase of a complex number, respectively. The operator $[x]^{+}$ is defined as $[x]^{+} \triangleq \max\{x,0\}$. The symbol $\mathbf{S} \succeq \mathbf{0}$ indicates that $\mathbf{S}$ is a positive semidefinite matrix.  The symbol $\mathcal{CN}\left(\mathbf{0}, \boldsymbol{\Sigma}\right)$ denotes the distribution of a circularly symmetric complex Gaussian vector with mean $\mathbf{0}$ and covariance matrix $\boldsymbol{\Sigma}$. The symbol $\mathbf{I}_M$ represents the identity matrix of size $M\times M$. Finally, the symbol $e$ represents the Euler number, and $j = \sqrt{-1}$ denotes the imaginary unit. 

\section{System Model And Problem Formulation}
\label{Sec-SysMod}
In this section, we describe the PLS model of the active IRS-assisted MIMO-OFDM SWIPT system and then formulate the secrecy sum-rate maximization problem.

\subsection{System Model}
\begin{figure}[t]
	\centering
	\includegraphics[width=0.5\textwidth]{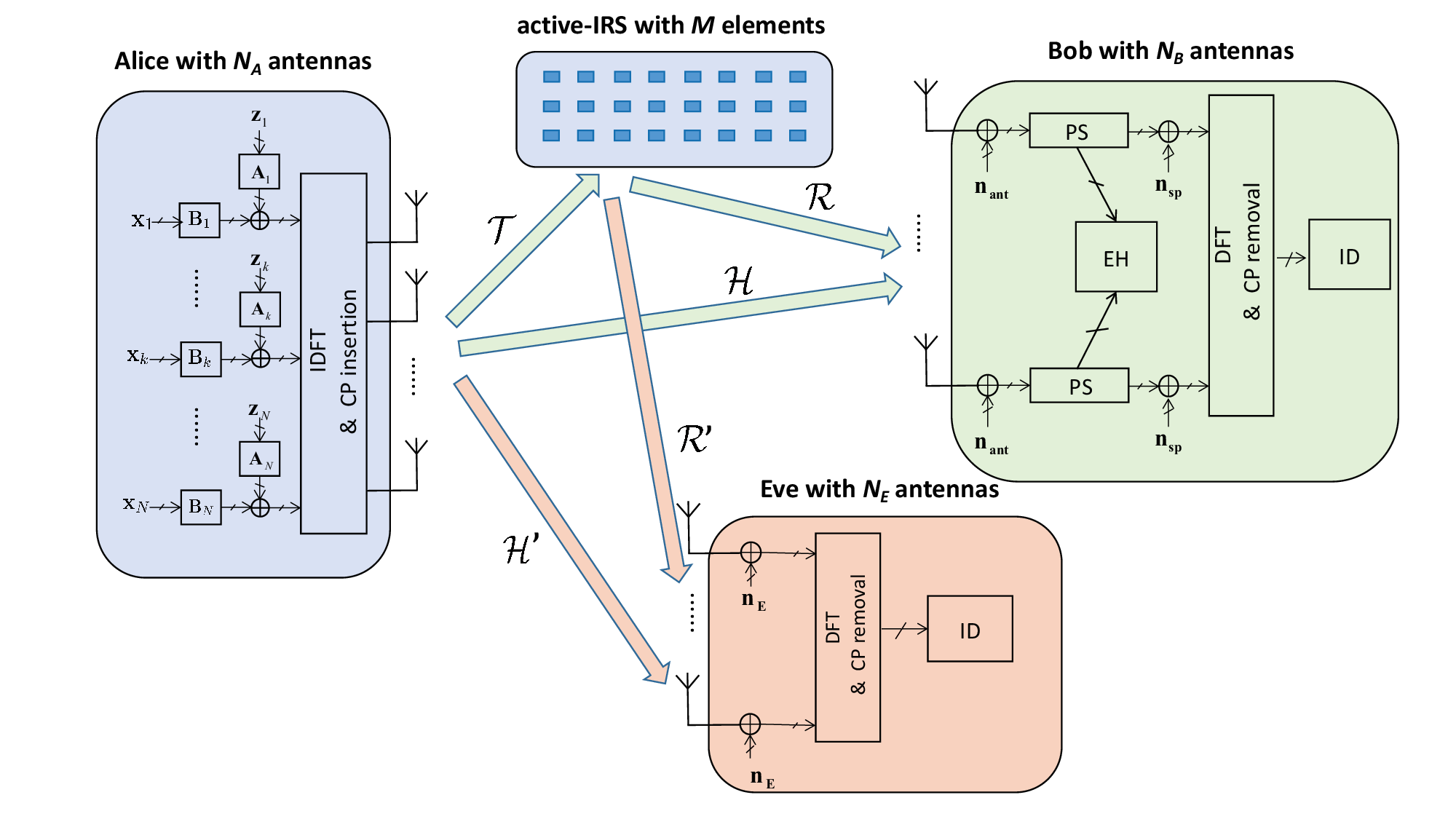}
	\caption{Schematic diagram of a MIMO-OFDM SWIPT system, where Alice sends information to Bob in the presence of an active IRS and an eavesdropper Eve.}
	\label{Sys_Diag}
\end{figure}

As shown in Fig. \ref{Sys_Diag}, we consider an active IRS-assisted SWIPT secrecy system, which comprises a source node (Alice) with $N_{\rm A}$ antennas, an active IRS with $M$ reflecting elements, a PS-based legitimate node (Bob) with $N_{\rm B}$ antennas, and an eavesdropping node (Eve) with $N_{\rm E}$ antennas. To combat the frequency selectivity of the channel, we adopt OFDM modulation in the system, dividing the total bandwidth into $N$ subcarriers. We define $N_{\rm s} \triangleq \min\{N_{\rm A}, N_{\rm B}\}$ as the number of spatial multiplexing layers on each subcarrier. Next, we describe the system's five critical signal processing steps, starting with precoding at Alice.

\subsubsection{\underline{Precoding at Alice}}
Let $\mathbf{x}_k \sim \mathcal{CN}\left(\mathbf{0}, \mathbf{I}_{N_{\rm s}}\right)$ and $\mathbf{z}_k \sim \mathcal{C N}\left(\mathbf{0}, \mathbf{I}_{N_{\rm A}}\right)$ denote the information and AN symbol vectors on subcarrier $k$, respectively; $\mathbf{B}_k \in \mathbb{C}^{N_{\rm A} \times N_{\rm s}}$ and $\mathbf{A}_k \in \mathbb{C}^{N_{\rm A} \times N_{\rm A}}$ represent the corresponding information and AN precoding matrices. Then, the frequency-domain vector to be transmitted on each subcarrier $k$ is expressed as 
\begin{align}
	\mathbf{s}_k = \mathbf{B}_k\mathbf{x}_k + \mathbf{A}_k\mathbf{z}_k, \,\, 1 \leq k \leq N.
\end{align}
By stacking $\{\mathbf{s}_k\}_{k=1}^N$ into a vector, i.e., $\mathbf{s} \triangleq [\mathbf{s}_1^{ T}, \cdots, \mathbf{s}_N^{ T}]^{T}$, and performing the inverse discrete Fourier transform (IDFT), the resulting time-domain signal is given by
\begin{align}
	\tilde{\mathbf{s}} & = \left(\mathbf{F}^{H} \otimes \mathbf{I}_{N_{\rm A}}\right) \mathbf{s} = \left(\mathbf{F}^{H} \otimes \mathbf{I}_{N_{\rm A}}\right) \left(\mathbf{B}\mathbf{x} + \mathbf{A}\mathbf{z}\right),
\end{align}
where $\mathbf{F}$ denotes the $N \times N$ normalized Fourier matrix, of which the $(m,n)$-th element is given by $(1/\sqrt{N}) e^{-j2\pi mn/N}$, $\mathbf{B} \triangleq \operatorname{blkdiag}\left(\{\mathbf{B}_k\}_{k=1}^N\right)$, $\mathbf{x} \triangleq [\mathbf{x}_1^{ T}, \cdots, \mathbf{x}_N^{ T}]^{T}$, $\mathbf{A} \triangleq \operatorname{blkdiag}\left(\{\mathbf{A}_k\}_{k=1}^N\right)$, and $\mathbf{z} \triangleq [\mathbf{z}_1^{ T}, \cdots, \mathbf{z}_N^{ T}]^{T}$. After appending a cyclic prefix (CP) of $N_{\rm A}N_{\rm cp}$ samples to $\tilde{\mathbf{s}}$, the time-domain signal is transmitted through $N_{\rm A}$ RF chains.

\subsubsection{\underline{Power Splitting at Bob}} 
The time-domain channel for the Alice-Bob link is modeled by a finite-duration impulse response $\{\boldsymbol{\mathcal{H}}_l \in \mathbb{C}^{N_{\rm B}\times N_{\rm A}}\}_{l=0}^{L_{\rm AB}-1}$, where $L_{\rm AB}$ is the number of delay taps, and $\boldsymbol{\mathcal{H}}_l$ is the channel impulse response (CIR) matrix associated with the $l$-th tap. Similarly, the time-domain channels for the Alice-IRS and IRS-Bob links are modeled by $\{\boldsymbol{\mathcal{T}}_l \in \mathbb{C}^{M \times N_{\rm A}}\}_{l=0}^{L_{\rm AI}-1}$ and $\{\boldsymbol{\mathcal{R}}_l \in \mathbb{C}^{N_{\rm B}\times M}\}_{l=0}^{L_{\rm IB}-1}$, respectively. We assume that the length of CP is larger than the maximum power delay spread between Alice and Bob, i.e., $N_{\rm cp} \geq \max\{L_{\rm AB}, L_{\rm AI} + L_{\rm IB} - 1\}$. Then, the received time-domain signal after CP removal at Bob is expressed as
\begin{align}
	\tilde{\mathbf{y}}_{\rm B} & = \left(\boldsymbol{\mathcal{H}} + \boldsymbol{\mathcal{R}} (\mathbf{I}_N \otimes \boldsymbol{\Phi}) \boldsymbol{\mathcal{T}}\right) \tilde{\mathbf{s}} + \boldsymbol{\mathcal{R}} (\mathbf{I}_N \otimes \boldsymbol{\Phi})\tilde{\mathbf{n}}_{\rm I} + \tilde{\mathbf{n}}_{\rm ant}, \label{rs}
\end{align}
where $\boldsymbol{\mathcal{H}} \triangleq \operatorname{blkcirc}\left(\boldsymbol{\mathcal{H}}_0, \cdots, \boldsymbol{\mathcal{H}}_{L_{\rm AB}-1}, \cdots, \mathbf{0}_{N_{\rm B} \times N_{\rm A}}\right)$ is a block circulant matrix of size ${NN_{\rm B}  \times NN_{\rm A}}$. Likewise, $\boldsymbol{\mathcal{T}} \in \mathbb{C}^{NM \times NN_{\rm A}}$ and $\boldsymbol{\mathcal{R}} \in \mathbb{C}^{NN_{\rm B} \times NM}$ are block circulant matrices defined by $\{\boldsymbol{\mathcal{T}}_l\}_{l=0}^{L_{\rm TI} - 1}$ and $\{\boldsymbol{\mathcal{R}}_l\}_{l = 0}^{L_{\rm IR} - 1}$, respectively. $\boldsymbol{\Phi}=\operatorname{diag}\left(\{\alpha_m e^{j \psi_m}\}_{m=1}^M\right)$ denotes the reflection-coefficient matrix of the active-IRS, in which $\alpha_m \geq 0,\,\forall m$, and $\psi_m \in [0,2\pi),\, \forall m$, are the amplitude and the phase shift of the $m$-th reflecting element, respectively. Also, $\tilde{\mathbf{n}}_{\rm I} \sim \mathcal{CN}\left(\mathbf{0}, \sigma_{\mathrm{I}}^2\mathbf{I}_{NM}\right)$ and $\tilde{\mathbf{n}}_{\rm ant} \sim \mathcal{CN}\left(\mathbf{0}, \sigma_{\mathrm{ant}}^2\mathbf{I}_{NN_{\rm B}}\right)$ are the cyclic symmetric additive white Gaussian noises (AWGN) introduced by the active IRS and Bob's antennas, respectively. In contrast to the passive IRS, the noise introduced at the active IRS is non-negligible and would be amplified by the reflecting elements, as shown in \eqref{rs}. Bob adopts the PS scheme to extract information and energy from the received signal, splitting the received signal power into two parts for information decoding (ID) and energy harvesting (EH). Accordingly, the time-domain signals dedicated for ID and EH are respectively given by
\begin{align}
	\tilde{\mathbf{y}}_{\rm B}^{\rm ID} &= \left(\mathbf{I}_{N}\otimes\mathbf{D}^{1/2}\right) \tilde{\mathbf{y}}_{\rm B} + \tilde{\mathbf{n}}_{\rm sp},\\
	\tilde{\mathbf{y}}_{\rm B}^{\rm EH} &= \left(\mathbf{I}_{N}\otimes\left(\mathbf{I}_{N_{\rm B}} - \mathbf{D}\right)^{1/2}\right) \tilde{\mathbf{y}}_{\rm B}, \label{ehsig}
\end{align}
where $\mathbf{D} \triangleq \operatorname{diag}\left(\{d_r\}_{r=1}^{N_{\rm B}}\right)$, and $d_r \in [0,1]$ represents the PS factor at the $r$-th antenna of Bob, and $\tilde{\mathbf{n}}_{\rm sp} \sim \mathcal{CN}\left(\mathbf{0}, \sigma_{\rm sp}^2\mathbf{I}_{NN_{\rm B}}\right)$ denotes the AWGN vector introduced during the signal processing of the ID branch.

\subsubsection{\underline{Information Decoding at Bob}}
At the ID branch of Bob, the time-domain signal is converted to the frequency domain by performing the DFT operation, i.e.,
\begin{align}
	\mathbf{y}_{\rm B}^{\rm ID} &=\left(\mathbf{F} \otimes \mathbf{I}_{N_{\rm B}}\right)\tilde{\mathbf{y}}_{\rm B}^{\rm ID}  \label{id1}\\
	 &= \left(\mathbf{I}_{N}\otimes\mathbf{D}^{1/2}\right) \Big(\left(\mathbf{H} + \mathbf{R} \left(\mathbf{I}_N \otimes \boldsymbol{\Phi}\right) \mathbf{T}\right) \left(\mathbf{B}\mathbf{x} + \mathbf{A}\mathbf{z}\right) \notag \\
	&\quad {} + \mathbf{R} \left(\mathbf{I}_N \otimes \boldsymbol{\Phi}\right)\mathbf{n}_{\rm I} + \mathbf{n}_{\rm ant} \Big) + \mathbf{n}_{\rm sp}, \label{id2}
\end{align}
where $\mathbf{H} \triangleq (\mathbf{F} \otimes \mathbf{I}_{N_{\rm B}}) \boldsymbol{\mathcal{H}} (\mathbf{F}^{ H}\otimes \mathbf{I}_{N_{\rm A}})$, $\mathbf{R} \triangleq (\mathbf{F}\otimes \mathbf{I}_{N_{\rm B}}) \boldsymbol{\mathcal{R}} (\mathbf{F}^{ H} \otimes \mathbf{I}_{M})$,  $\mathbf{T} \triangleq (\mathbf{F} \otimes \mathbf{I}_{M}) \boldsymbol{\mathcal{T}} (\mathbf{F}^{ H} \otimes \mathbf{I}_{N_{\rm A}})$, $\mathbf{n}_{\rm I} \triangleq \left(\mathbf{F} \otimes \mathbf{I}_{M}\right) \tilde{\mathbf{n}}_{\rm I}$, $\mathbf{n}_{\rm ant} \triangleq \left(\mathbf{F} \otimes \mathbf{I}_{N_{\rm B}}\right) \tilde{\mathbf{n}}_{\rm ant}$, and $\mathbf{n}_{\rm sp} \triangleq \left(\mathbf{F} \otimes \mathbf{I}_{N_{\rm B}}\right) \tilde{\mathbf{n}}_{\rm sp}$. Using the property of block circulant matrices, we can express $\mathbf{H} = \operatorname{blkdiag}\left(\{\mathbf{H}_k\}_{k=1}^N\right)$, $\mathbf{R} = \operatorname{blkdiag} \left(\{\mathbf{R}_k\}_{k=1}^N\right)$, $\mathbf{T} = \operatorname{blkdiag}\left(\{\mathbf{T}_k\}_{k=1}^N\right)$, in which $\mathbf{H}_k \in \mathbb{C}^{N_{\rm B} \times N_{\rm A}}$, $\mathbf{R}_k \in \mathbb{C}^{N_{\rm R} \times M}$, and $\mathbf{T}_k \in \mathbb{C}^{M \times N_{\rm A}}$ are the channel frequency response (CFR) on subcarrier $k$ for the Alice-Bob, IRS-Bob, and Alice-IRS links, respectively. Based on \eqref{id2}, the received frequency-domain signals on each subcarrier $k$ is given by
\begin{align}
	\mathbf{y}_{{\rm B},k}^{\rm ID}  &= \mathbf{D}^{1/2} \bar{\mathbf{H}}_k\mathbf{B}_k\mathbf{x}_k \notag  + \mathbf{D}^{1/2} \\
	&\quad \times \left(\bar{\mathbf{H}}_k\mathbf{A}_k\mathbf{z}_k + \mathbf{R}_k\mathbf{\Phi}\mathbf{n}_{{\rm I},k} +\mathbf{n}_{{\rm ant},k}\right) + \mathbf{n}_{{\rm sp},k}, \,\,\forall k \label{idsign} 
\end{align}
where $\mathbf{\bar{H}}_k \triangleq \mathbf{H}_k + \mathbf{R}_k \boldsymbol{\Phi} \mathbf{T}_k$ denotes the effective CFR between Alice and Bob on subcarrier $k$; $\mathbf{n}_{{\rm I}, k}$, $\mathbf{n}_{{\rm ant},k}$ and $\mathbf{n}_{{\rm sp}, k}$ are the sub-vectors formed by taking the $((k-1) N + 1)$-th to the $(k N)$-th entries from $\mathbf{n}_{\rm I}$, $\mathbf{n}_{\rm ant}$ and $\mathbf{n}_{\rm sp}$, respectively. With (\ref{idsign}), the achievable data rate on each subcarrier $k$ at the ID branch of Bob can be computed as
\begin{align}
	\label{ratel}
	R_{{\rm B},k} = \operatorname{log}_2 \operatorname{det} \left( \mathbf{I}_{N_{\rm B}} + \bar{\mathbf{H}}_k\mathbf{B}_k\mathbf{B}_k^H\bar{\mathbf{H}}_k^H\mathbf{\Gamma}_k^{-1}\right), \,\, \forall k 
\end{align}
where $\mathbf{\Gamma}_k \triangleq \bar{\mathbf{H}}_k\mathbf{A}_k\mathbf{A}_k^H\bar{\mathbf{H}}_k^H + \sigma_{\rm I}^2  \mathbf{R}_k\mathbf{\Phi}\mathbf{\Phi}^H\mathbf{R}_k^H + \sigma_{\mathrm{ant}}^2 \mathbf{I}_{N_{\rm B}} + \sigma_{\mathrm{sp}}^2 \mathbf{D}^{-1}$ represents the covariance matrix of the effective noise on subcarrier $k$ at the ID branch of Bob.

\subsubsection{\underline{Energy Harvesting at Bob}}
At the EH branch of Bob, the input RF power can be derived as $E_{\rm in} = \mathbb{E} [(\tilde{\mathbf{y}}_{\rm B}^{\rm EH})^H\tilde{\mathbf{y}}_{\rm B}^{\rm EH}] $. According to the Parseval's theorem and by performing similar manipulations as in \eqref{id1}-\eqref{idsign}, it can be further expressed as
\begin{align}
	E_{\rm in} &= \sum_{k = 1}^{N} \operatorname{Tr}\Big( \left(\mathbf{I} - \mathbf{D}\right) \big( \bar{\mathbf{H}}_k\left(\mathbf{B}_k\mathbf{B}_k^H + \mathbf{A}_k\mathbf{A}_k^H\right)\bar{\mathbf{H}}_k^H \notag \\ 
	& \hspace{3em} {}+  \sigma_{\rm I}^2  \mathbf{R}_k\mathbf{\Phi}\mathbf{\Phi}^H\mathbf{R}_k^H + \sigma_{\rm ant}^2\mathbf{I}_{N_{\rm B}}\big)\Big), \label{eh112}
\end{align}
Since the EH circuits contain nonlinear electronic components such as diodes, the conversion efficiency of the energy harvester varies with the input power. Therefore, Bob introduces a parametric nonlinear EH model to characterize the relationship between the harvested power and the input power \cite{8060616,8097000},
\begin{equation}
	\label{eh2}
	\gamma\left(E_{\rm in}\right) = \left[ \frac{E_{\rm m}}{e^{-\xi E_{\rm 0} + \nu}}\left(\frac{1 + e^{-\xi E_{\rm 0} + \nu}}{1 + e^{-\xi E_{\rm in} + \nu}} - 1\right) \right]^{+},
\end{equation}
where $\xi$ and $\nu$ are system parameters related to the circuit characteristics and $E_{\rm 0}$ and $E_{\rm m}$ denote the EH circuit's activation and saturation power levels, respectively \cite{8060616}.

\subsubsection{\underline{Information Eavesdropping at Eve}}
The time-domain channels for the Alice-Eve and IRS-Eve links are modeled by finite-duration impulse responses $\{\boldsymbol{\mathcal{H}}^{\prime}_l \in \mathbb{C}^{N_{\rm E} \times N_{\rm A}}\}_{l=0}^{L_{\rm AE}-1}$ and $\{\boldsymbol{\mathcal{R}}^{\prime}_l \in \mathbb{C}^{N_{\rm E}\times M}\}_{l=0}^{L_{\rm IE}-1}$, respectively. We assume that $N_{\rm cp} \geq \max\{L_{\rm AE}, L_{\rm AI} + L_{\rm IE} - 1\}$. To avoid inter-block interference at the legitimate user, Alice usually assigns a fairly large value to $N_{\rm cp}$ to cover most channels. Therefore, this requirement is not intended for Eve but rather for Bob. As such, the received time-domain signal after CP removal at Eve is expressed as
\begin{align}
\hspace{-0.25em}	\tilde{\mathbf{y}}_{\rm E} = (\boldsymbol{\mathcal{H}}^{\prime} + \boldsymbol{\mathcal{R}}^{\prime} (\mathbf{I}_N \otimes \boldsymbol{\Phi}) \boldsymbol{\mathcal{T}}) \tilde{\mathbf{s}} + \boldsymbol{\mathcal{R}}^{\prime} (\mathbf{I}_N \otimes \boldsymbol{\Phi})\tilde{\mathbf{n}}_{\rm I} + \tilde{\mathbf{n}}_{\rm E},
\end{align}
where $\boldsymbol{\mathcal{H}}^{\prime} \in \mathbb{C}^{NN_{\rm E} \times NN_{\rm A}}$ and $\boldsymbol{\mathcal{R}}^{\prime} \in \mathbb{C}^{NN_{\rm E} \times NM}$ are block circulant matrices defined by $\{\boldsymbol{\mathcal{H}}^{\prime}_l\}_{l=0}^{L_{\rm AE}-1}$ and $\{\boldsymbol{\mathcal{R}}^{\prime}_l \}_{l=0}^{L_{\rm IE}-1}$, respectively, and $\tilde{\mathbf{n}}_{\rm E} \sim \mathcal{CN}\left(\mathbf{0}, \sigma_{\mathrm{E}}^2\mathbf{I}_{NN_{\rm E}}\right)$ denotes the AWGN introduced by Eve's antennas. By performing similar manipulations as in \eqref{id1}-\eqref{idsign}, the received frequency-domain signals on each subcarrier $k$ are given by
\begin{align}
	\mathbf{y}_{{\rm E},k} =\bar{\mathbf{H}}_k^{\prime}\mathbf{B}_k\mathbf{x}_k + \bar{\mathbf{H}}_k^{\prime}\mathbf{A}_k\mathbf{z}_k + \mathbf{R}_k^{\prime}\mathbf{\Phi}\mathbf{n}_{{\rm I},k} + \mathbf{n}_{{\rm E},k}, \, \, \forall k \label{RE}
\end{align}
where $\bar{\mathbf{H}}_k^{\prime} \triangleq \mathbf{H}_k^{\prime} + \mathbf{R}_k^{\prime}\mathbf{\Phi}\mathbf{T}_k$ denotes the effective CFR between Alice and Eve on subcarrier $k$, and $\mathbf{H}_k^{\prime}$ and $\mathbf{R}_k^{\prime}$ are the $k$-th diagonal blocks of $(\mathbf{F} \otimes \mathbf{I}_{N_{\rm E}}) \boldsymbol{\mathcal{H}}^{\prime} (\mathbf{F}^{ H}\otimes \mathbf{I}_{N_{\rm A}})$ and $(\mathbf{F} \otimes \mathbf{I}_{N_{\rm E}}) \boldsymbol{\mathcal{R}}^{\prime} (\mathbf{F}^{ H}\otimes \mathbf{I}_{M})$, respectively. Also, $\mathbf{n}_{{\rm E}, k}$ is the vector comprising the $((k-1) N + 1)$-th to the $(k N)$-th entries of $\mathbf{n}_{\rm E}$. With (\ref{RE}), the eavesdropping data rate on each subcarrier $k$ at Eve can be computed as
\begin{align}
	\label{ratee}
	R_{{\rm E},k} = \operatorname{log}_2 \operatorname{det} \left( \mathbf{I}_{N_{\rm E}} + \bar{\mathbf{H}}_k^{\prime}\mathbf{B}_k\mathbf{B}_k^H\bar{\mathbf{H}}_k^{\prime H}\mathbf{\Gamma}_k^{\prime -1}\right), \,\, \forall k
\end{align}
where $\mathbf{\Gamma}_k^{\prime} \triangleq \bar{\mathbf{H}}_k^{\prime}\mathbf{A}_k\mathbf{A}_k^H\bar{\mathbf{H}}_k^{\prime H} + \sigma_{\rm I}^2  \mathbf{R}_k^{\prime}\mathbf{\Phi}\mathbf{\Phi}^H\mathbf{R}_k^{\prime H}  + \sigma_{\rm E}^2\mathbf{I}_{N_{\rm E}}$ is the covariance matrix of the effective noise on subcarrier $k$.

It is evident from (\ref{ratel}) and (\ref{ratee}) that the noise introduced at the active IRS causes interference to both legitimate users and eavesdroppers. Therefore, it can be exploited as an AN to improve the system's secrecy sum rate by carefully designing the IRS matrix.

\begin{remark}
	\label{rmk2}
	To derive the relevant performance upper bounds for the considered system, we assume that the perfect channel state information (CSI) of all links is available at all nodes, like \cite{9903846,9913501}. The existing works \cite{8937491,9195133} have proposed efficient methods for channel estimation of IRS-assisted OFDM systems. Moreover, it is safe to assume that Eve's CSI is known when Eve is registered but untrusted by Bob, e.g., in broadcast scenarios with confidential messages.
\end{remark}

\begin{remark}
	\label{rmk22}
	If the eavesdropper is a registered user in the network, obtaining its CSI is typically feasible \cite{9913501}. For instance, in broadcast networks, the eavesdropper may be a registered user allowed to receive their own and common messages but not permitted to receive confidential messages intended for other users \cite{5550390}. Similarly, in unicast-multicast streaming systems, the base station can serve multiple users simultaneously but sometimes needs to send specific information to a certain user \cite{9855515}. Additionally, in Internet of Things (IoT) networks, the widespread deployment of devices and advanced network management tools, e.g., Aruba AirWave, enable the collection of CSIs from all participating entities, including potential eavesdroppers \cite{10535422}. 
\end{remark}

\subsection{Problem Formulation}	
As discussed above, the considered OFDM system can be regarded as the parallel of $N$ wiretap channels. With (\ref{ratel}) and (\ref{ratee}), the secrecy rate on each subcarrier $k$ is defined as
\begin{align}
	\label{secrate}
	R_{{\rm Sec},k} \triangleq \left[ R_{{\rm B},k} - R_{{\rm E},k}\right]^{+}, \,\, \forall k.
\end{align}
Since the optimal secrecy rate is usually non-negative, we omit the operator $[\cdot]^{+} $ in the rest of this paper.

This paper aims to maximize the secrecy sum rate over all subcarriers. Meanwhile, Alice's transmit power and the Active-IRS's reflecting power should be within their respective budgets. Besides, Bob should harvest enough energy at the EH branch to support its operation. As a result, the optimization problem is mathematically formulated as
\begin{align*}
	\mathcal{P} 1:  \max_{\{\mathbf{B}_k,\mathbf{A}_k\}_{k=1}^N,\mathbf{\Phi},\mathbf{D}} &\sum_{k = 1}^N R_{{\rm Sec},k} \\
	\mbox{s.t.} \quad  C_1: \,&\sum_{k=1}^{N} \operatorname{Tr}\left(\mathbf{B}_k \mathbf{B}_k^H + \mathbf{A}_k \mathbf{A}_k^H\right)  \leq P_{\rm A},\\
	C_2: \, &\sum_{k = 1}^N\operatorname{Tr}\big(\boldsymbol{\Phi}\mathbf{T}_k( \mathbf{B}_k\mathbf{B}_k^H + \mathbf{A}_k\mathbf{A}_k^H)\mathbf{T}_k^H\boldsymbol{\Phi}^H\big) \notag \\
	& \quad {}+ \sigma_{\rm I}^2 \operatorname{Tr}( \boldsymbol{\Phi}\boldsymbol{\Phi}^H ) \leq P_{\rm I},\\
	C_3: \, &\gamma\left(E_{\rm in}\right) \geq E_{\rm th},\\
	C_4: \, &0 \leq [\mathbf{D}]_{r,r} \leq 1, \,\,  \forall r
\end{align*}
where $P_{\rm A}$ denotes the maximum transmit power at Alice, $P_{\rm I}$ represents the maximum reflecting power at the active IRS, and $E_{\rm th}$ is the harvested power target at the EH branch of Bob. It can be seen that $\mathcal{P} 1$ is highly non-convex and challenging to solve since the matrix-based optimization variables are intricately coupled in the objective function, the reflecting power constraint, and the harvested power constraint. Moreover, in contrast to the passive IRS-assisted system, the active IRS introduces non-negligible noise during reflecting incident signals, which would be amplified by the IRS matrix, thus imposing another challenge for solving this problem.

\begin{remark}[Problem formulation for the passive-IRS case]
	\label{rmk3}
	Since the thermal noise introduced by the passive-IRS is negligible, the noise power $\sigma_{ \rm I}^2$ is set to zero in the achievable data rate at Bob \eqref{ratel}, the input RF power at Bob \eqref{eh112}, and the eavesdropping data rate at Eve \eqref{ratee}. Furthermore, due to the passive nature of the IRS, the reflecting power constraint $C_2$ in $\mathcal{P} 1$ is replaced by the uni-modulus constraints $\vert [\boldsymbol{\Phi}]_{m,m} \vert = 1, \,\, \forall m$.
\end{remark}

For ease of tractability, the main system parameters pertaining to the algorithm designs in the subsequent sections are summarized in Table \ref{table_freqSymbol}.

\begin{table}[t] \footnotesize 
	\renewcommand{\arraystretch}{1.1}
	\caption{List of Main System Parameters}
	\label{table_freqSymbol}
	\centering
	\begin{tabular}{c||l}
		\toprule[1.5pt]
		\textbf{Symbols}             					& \textbf{Descriptions} 		\\
		\hline 
		$N, N_{\rm A}, N_{\rm S}$						& \# of subcarriers/Tx antennas/spatial layers\\
		 
		$M, N_{\rm B}, N_{\rm E}$						& \# of reflecting elements/Rx antennas at Bob/Eve\\
		 
	    $\{\mathbf{B}_k, \mathbf{A}_k\}_{k=1}^N$	    & Information/AN precoding matrices at Alice\\
		 
		$\boldsymbol{\Phi}$	                & Reflection-coefficient matrix at the active-IRS\\
		 
		$ \mathbf{D}$	                & Power splitting matrix at Bob\\
		 
		$\{\mathbf{H}_k, \mathbf{T}_k, \mathbf{R}_k\}_{k=1}^N$	    & CFRs of the Alice-Bob/Alice-IRS/IRS-Bob link\\
		 
		$\{\mathbf{H}_k^{\prime}, \mathbf{R}_k^{\prime}\}_{k=1}^N$	& CFRs of the Alice-Eve/IRS-Eve link\\
		 
		$\{\bar{\mathbf{H}}_k, \bar{\mathbf{H}}_k^{\prime}\}_{k=1}^N$	& Effective CFRs between Alice and Bob/Eve\\
		 
		$\sigma_{ \rm I}^2,  \sigma_{ \rm E}^2$   & Noise power at the active-IRS/Eve  \\
		 
		$\sigma_{ \rm ant}^2, \sigma_{ \rm sp}^2$  & Noise power of Rx antenna/ID branch at Bob  \\
		
	    $\xi, \nu, E_{\rm 0}, E_{\rm m}$                         & Non-linear EH model parameters	\\
		
		$E_{\rm th}$  & The minimum EH requirement at Bob\\
		 
		$P_{\rm A}, P_{\rm I}$  & Power budgets at Alice/active-IRS\\
		\bottomrule[1.5pt]  
	\end{tabular}
\end{table}

\section{BCD-Based Algorithm} 
\label{Sec-BCD}
In this section, we first reformulate $\mathcal{P} 1$ into a more tractable form and then develop a BCD-based algorithm to solve it. Within the BCD framework, the first-order optimality condition is used to obtain the optimal auxiliary variables; the SDR method is adopted to optimize the information precoding matrices; the penalty method is applied to derive the optimal PS ratios; and the IA method is employed to optimize the IRS matrix. Finally, the convergence and complexity of the BCD-based algorithm are analyzed.

\subsection{Reformulation of the Original Problem}
It is clear that the EH constraint $C_3$ in $\mathcal{P} 1$ is intractable since the nonlinear EH model shown in \eqref{eh2} is a sophisticated function of  $\{\mathbf{B}_k,\mathbf{A}_k\}_{k=1}^N,\mathbf{\Phi}$, and $\mathbf{D}$. By combining (\ref{eh112}) and (\ref{eh2}), the original EH constraint $C_3$ can be equivalently transformed into a more tractable form as follows:
\begin{equation}
	C_5:\,\, E_{\rm in} \geq \bar{\gamma}(E_{\rm th}),
\end{equation}
where
\begin{equation}
	\bar{\gamma}(x) = \begin{cases}+\infty, & \text { if } x \geq E_{\max } \\ 
		E_{\rm 0} - \xi^{-1} \operatorname{ln}\Big(\frac{E_{m} - x}{E_{\rm m} + x e^{-\xi E_{\rm 0} + \nu}}\Big), & \text { if } 0<x<E_{\max } \\ 
		0, & \text { if } x \leq 0\end{cases}
\end{equation}
is the pseudo-inverse of the nonlinear EH function $\gamma(x)$.

Next, we reformulate the objective function of $\mathcal{P}1$ by using the relationship between the achievable data rate and the weighted minimum mean-square error (WMMSE) \cite{5756489}. Upon rearrangements, the secrecy rate on each subcarrier \eqref{secrate} can be rewritten as \eqref{ms}, shown at the top of the page,
\begin{figure*}[!t]
	\setlength{\arraycolsep}{0.0em}
	\begin{align}
			R_{{\rm Sec},k} & = \operatorname{log}_2 \operatorname{det} \left( \mathbf{I}_{N_{\rm B}} + \bar{\mathbf{H}}_k\mathbf{B}_k\mathbf{B}_k^H\bar{\mathbf{H}}_k^H\mathbf{\Gamma}_k^{-1}\right) - \operatorname{log}_2 \operatorname{det} \left(\mathbf{I}_{N_{\rm E}} + \bar{\mathbf{H}}_k^{\prime}\mathbf{B}_k\mathbf{B}_k^H\bar{\mathbf{H}}_k^{\prime H} \mathbf{\Gamma}_k^{\prime -1}\right) \notag \\
			& = \operatorname{log}_2 \operatorname{det} \left( \mathbf{I}_{N_{\rm B}} + \bar{\mathbf{H}}_k\mathbf{B}_k\mathbf{B}_k^H\bar{\mathbf{H}}_k^H\mathbf{\Gamma}_k^{-1}\right)  +  \operatorname{log}_2 \operatorname{det} \left( \mathbf{\Gamma}_k^{\prime} \right) + \operatorname{log}_2 \operatorname{det} \left( \mathbf{\Gamma}_k^{\prime} + \bar{\mathbf{H}}_k^{\prime}\mathbf{B}_k\mathbf{B}_k^H\bar{\mathbf{H}}_k^{\prime H}\right)^{-1} \notag \\
			& = \underbrace{ \operatorname{log}_2 \operatorname{det} \left( \mathbf{I}_{N_{\rm B}} + \bar{\mathbf{H}}_k\mathbf{B}_k\mathbf{B}_k^H\bar{\mathbf{H}}_k^H\mathbf{\Gamma}_k^{-1}\right)}_{\mathcal{I}_{1,k}}  + \underbrace{ \operatorname{log}_2 \operatorname{det} \left( \mathbf{I}_{N_{\rm E}} + \bar{\mathbf{H}}_k^{\prime}\mathbf{A}_k\mathbf{A}_k^H\bar{\mathbf{H}}_k^{\prime H}\bar{\mathbf{\Gamma}}_k^{\prime -1} \right)}_{\mathcal{I}_{2,k}}\notag \\
			& \quad + \underbrace{\operatorname{log}_2 \operatorname{det} \Big( \mathbf{I}_{N_{\rm E}} + \frac{\sigma_{\rm I}^2}{\sigma_{\rm E}^2}  \mathbf{R}_k^{\prime}\mathbf{\Phi}\mathbf{\Phi}^H\mathbf{R}_k^{\prime H} \Big)}_{\mathcal{I}_{3,k}} + \underbrace{ \operatorname{log}_2 \operatorname{det} \left( \mathbf{\Gamma}_k^{\prime} + \bar{\mathbf{H}}_k^{\prime}\mathbf{B}_k\mathbf{B}_k^H\bar{\mathbf{H}}_k^{\prime H}\right)^{-1}}_{\mathcal{I}_{4,k}} + \kappa_1, \,\, \forall k \label{ms}
	\end{align}
	\setlength{\arraycolsep}{5pt}
	\hrulefill
\end{figure*}
where $\bar{\mathbf{\Gamma}}_k^{\prime} \triangleq  \sigma_{\rm E}^2\mathbf{I}_{N_{\rm E}} + \sigma_{\rm I}^2  \mathbf{R}_k^{\prime}\mathbf{\Phi}\mathbf{\Phi}^H\mathbf{R}_k^{\prime H}$, and $\kappa_1 \triangleq N_{\rm E}\log_2(\sigma_{\rm E}^2)$. It is observed that $\mathcal{I}_{1,k}$, $\mathcal{I}_{2,k}$ and $\mathcal{I}_{3,k}$ in \eqref{ms} can be regarded as the achievable data rates of three virtual communication systems, where $\mathbf{B}_k$, $\mathbf{A}_k$, and $\boldsymbol{\Phi}$ denote their precoding matrices, $\bar{\mathbf{H}}_k$, $\bar{\mathbf{H}}_k^{\prime}$, and $\mathbf{R}_k^{\prime}$ denote their channel matrices, and $\mathbf{\Gamma}_k$, $\bar{\mathbf{\Gamma}}_k^{\prime}$, and $(\sigma_{ \rm E}^2/\sigma_{\mathrm{I}}^2)\mathbf{I}$ represent their covariance matrices of effective noises. According to \cite{5756489}, the achievable data rate of a system can be expressed as a function w.r.t. the WMMSE matrix of the recovered signals. As a result, we can rewrite $\mathcal{I}_{1,k}$, $\mathcal{I}_{2,k}$ and $\mathcal{I}_{3,k}$ into the following equivalent forms:
\begin{align} 
	\mathcal{I}_{1,k} & = \max_{\mathbf{W}_k \succeq \mathbf{0},\mathbf{C}_k} \,\, \operatorname{log}_2 \operatorname{det}(\mathbf{W}_k) - \operatorname{Tr} ( \mathbf{W}_k \mathbf{E}_k) + N_{\rm S},  \\
	\mathcal{I}_{2,k} & = \max_{\mathbf{W}_k^{\prime} \succeq \mathbf{0},\mathbf{C}_k^{\prime}} \,\, \operatorname{log}_2 \operatorname{det}( \mathbf{W}_k^{\prime} ) - \operatorname{Tr} ( \mathbf{W}_k^{\prime} \mathbf{E}_k^{\prime}) + N_{\rm A}, \\
	\mathcal{I}_{3,k} & = \max_{\bar{\mathbf{W}}_k^{\prime} \succeq \mathbf{0},\bar{\mathbf{C}}_k^{\prime}}  \,\, \operatorname{log}_2 \operatorname{det}( \bar{\mathbf{W}}_k^{\prime} ) - \operatorname{Tr} ( \bar{\mathbf{W}}_k^{\prime} \bar{\mathbf{E}}_k^{\prime} ) + N_{\rm A},
\end{align}
where $\mathbf{E}_k \triangleq  (\mathbf{I}_{N_{\rm S}} - \mathbf{C}_k^H\bar{\mathbf{H}}_k\mathbf{B}_k)(\mathbf{I}_{N_{\rm S}} - \mathbf{C}_k^H\bar{\mathbf{H}}_k\mathbf{B}_k)^H + \mathbf{C}_k^H \mathbf{\Gamma}_k \mathbf{C}_k$, $\mathbf{E}_k^{\prime} \triangleq (\mathbf{I}_{N_{\rm A}} - \mathbf{C}_k^{\prime H}\bar{\mathbf{H}}_k^{\prime}\mathbf{A}_k)(\mathbf{I}_{N_{\rm A}} - \mathbf{C}_k^{\prime H}\bar{\mathbf{H}}_k^{\prime}\mathbf{A}_k)^H + \mathbf{C}_k^{\prime H} \bar{\mathbf{\Gamma}}_k^{\prime} \mathbf{C}_k^{\prime}$, and $\bar{\mathbf{E}}_k^{\prime} \triangleq (\mathbf{I}_{M} - \bar{\mathbf{C}}_k^{\prime H}\mathbf{R}_k^{\prime}\mathbf{\Phi})(\mathbf{I}_{M} - \bar{\mathbf{C}}_k^{\prime H}\mathbf{R}_k^{\prime}\mathbf{\Phi})^H + (\sigma_{\rm E}^2/\sigma_{\rm I}^2) \bar{\mathbf{C}}_k^{\prime H} \bar{\mathbf{C}}_k^{\prime}$. Note that $\mathbf{E}_k$, $\mathbf{E}_k^{\prime}$, and $\bar{\mathbf{E}}_k^{\prime}$ are, in essence, the MSE matrices of the virtual communication systems $\mathcal{I}_{1,k}$, $\mathcal{I}_{2,k}$, and $\mathcal{I}_{3,k}$, respectively; $\mathbf{W}_k$, $\mathbf{W}_k^{\prime}$, and $\bar{\mathbf{W}}_k^{\prime}$ are the corresponding weight matrices; and $\mathbf{C}_k$, $\mathbf{C}_k^{\prime}$, and $\bar{\mathbf{C}}_k^{\prime}$ are the corresponding equalization matrices. Since the rate-WMMSE relationship cannot be applied to handle the last term $\mathcal{I}_{4,k}$ in (\ref{ms}), we resort to the following lemma.

\begin{lemma} \label{Lemma-1}
	Given $\mathbf{E} \succ 0 \in \mathbb{C}^{N \times N}$, then we have
	\begin{align}
		\max _{\mathbf{S} \succ 0 \in \mathbb{C}^{N \times N}} f(\mathbf{S})=\log_2 \operatorname{det}\left(\mathbf{E}\right)^{-1}, \label{lem1}
	\end{align}
	where $f(\mathbf{S}) \triangleq -\operatorname{Tr}(\mathbf{S E})+\log_2\det(\mathbf{S}) + N$.
\end{lemma}

\begin{IEEEproof}
	This lemma can be readily proved using the Fenchel conjugate arguments \cite[P.~49]{R2}.
\end{IEEEproof}

By virtue of Lemma~\ref{Lemma-1}, the last term $\mathcal{I}_{4,k}$ in (\ref{ms}) can be equivalently rewritten as
\begin{align}
	\mathcal{I}_{4,k}  = \max_{\{\mathbf{W}_k^{\prime \prime} \succeq \mathbf{0}\}} \operatorname{log}_2 \operatorname{det}( \mathbf{W}_k^{\prime \prime} )  - \operatorname{Tr} ( \mathbf{W}_k^{\prime \prime} \mathbf{E}_k^{\prime \prime}) + N_{\rm E},
\end{align}
with $\mathbf{E}_k^{\prime \prime} \triangleq \boldsymbol{\Gamma}_k^{\prime} + \bar{\mathbf{H}}_k^{\prime}\mathbf{B}_k\mathbf{B}_k^H\bar{\mathbf{H}}_k^{\prime H}$.

Based on the above analysis and dropping the constant terms, the original $\mathcal{P}1$ can be equivalently reformulated as
\begin{align*}
	\mathcal{P} 2:  \, \max_{\substack{ \mathbf{\Xi}, \boldsymbol{\Phi}, \mathbf{D} \\ \{\mathbf{B}_k,\mathbf{A}_k\}_{k=1}^N}}  \, & \sum_{k=1}^{N} \Big(\operatorname{log}_2 \operatorname{det}\left( \mathbf{W}_k \right) + \operatorname{log}_2 \operatorname{det}\left( \mathbf{W}_k^{\prime} \right) \notag \\
	& \quad   + \operatorname{log}_2 \operatorname{det}\left( \bar{\mathbf{W}}_k^{\prime} \right) + \operatorname{log}_2 \operatorname{det}\left( \mathbf{W}_k^{\prime \prime} \right)  \\
	& \quad   - \operatorname{Tr} \left( \mathbf{W}_k \mathbf{E}_k\right)  - \operatorname{Tr} \left( \mathbf{W}_k^{\prime} \mathbf{E}_k^{\prime}\right)\\
	& \quad   - \operatorname{Tr} \left( \bar{\mathbf{W}}_k^{\prime}\bar{\mathbf{E}}_k^{\prime}\right) - \operatorname{Tr}\left( \mathbf{W}_k^{\prime \prime}\mathbf{E}_k^{\prime \prime}\right)\Big) \\
	\mbox{s.t.} \quad \quad \quad &C_1, \, \, C_2, \, \, C_4, \, \, C_5, 
\end{align*}
where $\mathbf{\Xi} \triangleq \{ \mathbf{W}_k \succeq \mathbf{0}, \,\,  \mathbf{C}_k, \,\, \mathbf{W}_k^{\prime} \succeq \mathbf{0}, \,\,  \mathbf{C}_k^{\prime}, \,\, \bar{\mathbf{W}}_k^{\prime} \succeq \mathbf{0}, \,\,  \bar{\mathbf{C}}_k^{\prime}, \,\, \mathbf{W}_k^{\prime \prime} \succeq \mathbf{0} \}_{k=1}^N$. Although $\mathcal{P}2$ involves more auxiliary optimization variables than the original problem $\mathcal{P}1$, its objective function becomes much easier to handle, which facilitates the development of a BCD-based algorithm for solving this problem. Specifically, we can partition the optimization variables into four coordinate blocks, i.e., $\mathbf{\Xi}$, $\{\mathbf{B}_k,\mathbf{A}_k\}_{k=1}^N$, $\mathbf{D}$, and $\boldsymbol{\Phi}$. At each iteration, we optimize one block with other fixed coordinate blocks\footnote{In comparison with previous studies on maximizing the sum rate of IRS-assisted systems, such as \cite{9998527} and \cite{10553296}, the optimization of these components should not only maximize the achievable data rate for the legitimate user but also minimize information leakage to eavesdroppers. Furthermore, the optimization of the components $\{\mathbf{B}_k, \mathbf{A}_k\}_{k=1}^{N}$ and $\mathbf{\Phi}$ is non-convex due to the EH constraint, and therefore cannot be solved using the Lagrangian multiplier method as in previous works.}. In the following, we discuss the optimization of each block in detail.

\subsection{Optimization of the Auxiliary Variables $\mathbf{\Xi}$}
Given blocks $\{\mathbf{B}_k,\mathbf{A}_k\}_{k=1}^N$, $\mathbf{D}$, and $\boldsymbol{\Phi}$, $\mathcal{P} 2$ reduces to an unconstrained optimization problem. By setting the first-order derivative of the objective function w.r.t. $\mathbf{C}_k$, $\mathbf{C}_k^{\prime}$, $\bar{\mathbf{C}}_k^{\prime}$, 
$\mathbf{W}_k$, $\mathbf{W}_k^{\prime}$, $\bar{\mathbf{W}}_k^{\prime}$, and $\mathbf{W}_k^{\prime \prime}$ to zero, respectively, the optimal solutions can be obtained as, $\forall k$,
\begin{align}
\mathbf{C}_k^{\star} & = \left(\bar{\mathbf{H}}_k\mathbf{B}_k\mathbf{B}_k^H\bar{\mathbf{H}}_k^H + \mathbf{\Gamma}_k\right)^{-1} \bar{\mathbf{H}}_k\mathbf{B}_k, \label{a1}\\
	\mathbf{C}_k^{\prime \star} & = \left(\bar{\mathbf{H}}_k^{\prime}\mathbf{A}_k\mathbf{A}_k^H\bar{\mathbf{H}}_k^{\prime H} + \bar{\mathbf{\Gamma}}_k^{\prime }\right)^{-1}\bar{\mathbf{H}}_k^{\prime}\mathbf{A}_k,\label{a2}\\
	\bar{\mathbf{C}}_k^{\prime \star} & = \Big(\mathbf{R}_k^{\prime}\mathbf{\Phi}\mathbf{\Phi}^H\mathbf{R}_k^{\prime H} + \frac{\sigma_{\rm E}^2}{\sigma_{\rm I}^2}\mathbf{I}_{N_{\rm E}} \Big)^{-1}\mathbf{R}_k^{\prime}\mathbf{\Phi}, \label{a3} \\
	\mathbf{W}_k^{\star} &= \mathbf{E}_k^{-1},   
	\mathbf{W}_k^{\prime \star} = \mathbf{E}_k^{\prime -1}, 
	\bar{\mathbf{W}}_k^{\prime \star}  = \bar{\mathbf{E}}_k^{\prime -1}, 
	\mathbf{W}_k^{\prime \prime \star}  = \mathbf{E}_k^{\prime \prime -1}. \label{wopt}
\end{align}
Note that $\mathbf{C}_k^{\star}$, $\mathbf{C}_k^{\prime \star}$, and $\bar{\mathbf{C}}_k^{\prime \star}$ are the well-known Wiener equalizers for the virtual communication systems $\mathcal{I}_{1,k}$, $\mathcal{I}_{2,k}$, and $\mathcal{I}_{3,k}$, respectively.

\subsection{Optimization of the Precoding Matrices $\{\mathbf{B}_k,\mathbf{A}_k\}_{k=1}^N$}	
Given blocks $\mathbf{\Xi}$, $\mathbf{D}$, and $\boldsymbol{\Phi}$, the optimization problem corresponding to information/AN precoding matrices $\{\mathbf{B}_k,\mathbf{A}_k\}_{k=1}^N$ is given by
\begin{align*}
	\mathcal{P} 3: \hspace{-1em} \min_{\{\mathbf{B}_k,\mathbf{A}_k\}_{k=1}^N} &\sum_{k=1}^{N} \big(  \operatorname{Tr} \left( \mathbf{W}_k \mathbf{E}_k\right) + \operatorname{Tr} \left( \mathbf{W}_k^{\prime} \mathbf{E}_k^{\prime}\right) + \operatorname{Tr} \left( \mathbf{W}_k^{\prime \prime} \mathbf{E}_k^{\prime \prime}\right) \big) \\
	\mbox{s.t.} \qquad &C_1, \, \, C_2, \, \, C_5.
\end{align*}
Upon rearrangements, it can be further expanded as
\begin{align*}
	\mathcal{P} 4:  \min_{\{\mathbf{B}_k,\mathbf{A}_k\}_{k=1}^N} & \sum_{k=1}^{N} \Big( \operatorname{Tr} ( \mathbf{W}_k\mathbf{C}_k^H\bar{\mathbf{H}}_k\mathbf{B}_k\mathbf{B}_k^H\bar{\mathbf{H}}_k^H\mathbf{C}_k) \\
	& \qquad  + \operatorname{Tr} \left( \mathbf{W}_k\mathbf{C}_k^H\bar{\mathbf{H}}_k\mathbf{A}_k\mathbf{A}_k^H\bar{\mathbf{H}}_k^H\mathbf{C}_k\right)\\
	& \qquad  -  2\Re\left\{\operatorname{Tr} \left( \mathbf{W}_k\mathbf{C}_k^H\bar{\mathbf{H}}_k\mathbf{B}_k\right) \right\}  \\
	& \qquad  + \operatorname{Tr} \left( \mathbf{W}_k^{\prime}\mathbf{C}_k^{\prime H}\bar{\mathbf{H}}_k^{\prime}\mathbf{A}_k\mathbf{A}_k^H\bar{\mathbf{H}}_k^{\prime H}\mathbf{C}_k^{\prime}\right) \\
	& \qquad  -  2\Re\left\{ \operatorname{Tr} \left( \mathbf{W}_k^{\prime}\mathbf{C}_k^{\prime H}\bar{\mathbf{H}}_k^{\prime}\mathbf{A}_k\right)\right\}\\
	& \qquad  + \operatorname{Tr} \left( \mathbf{W}_k^{\prime \prime}\bar{\mathbf{H}}_k^{\prime}\mathbf{B}_k\mathbf{B}_k^H\bar{\mathbf{H}}_k^{\prime H}\right) \\
	& \qquad + \operatorname{Tr} \left( \mathbf{W}_k^{\prime \prime}\bar{\mathbf{H}}_k^{\prime}\mathbf{A}_k\mathbf{A}_k^H\bar{\mathbf{H}}_k^{\prime H}\right) \Big) \\
	\mbox{s.t.} \quad \quad &C_1, \, \, \, C_2, \,\, C_5.
\end{align*}
Due to the EH constraint $C_5$, $\mathcal{P} 4$ becomes a non-convex inhomogeneous quadratically constrained quadratic program (QCQP), generally NP-hard. In the following, we solve this problem by using the SDR technique. To be specific, we introduce variables $\mathbf{X}_{k} \triangleq \mathbf{B}_{k} \mathbf{B}_{k}^{H}$ and $\mathbf{Z}_{k} \triangleq \mathbf{A}_{k} \mathbf{A}_{k}^{H}$, $\forall k$, and then relax them as $\mathbf{X}_{k} \succeq \mathbf{B}_{k} \mathbf{B}_{k}^{H}$ and $\mathbf{Z}_{k} \succeq \mathbf{A}_{k} \mathbf{A}_{k}^{H}$, $\forall k$, which, based on the Schur complement lemma \cite{Derinkuyu2006OnTS}, can be expressed into linear matrix inequality (LMI) forms as
\begin{equation}
	\left[\begin{array}{cc}\mathbf{X}_{k} & \mathbf{B}_{k} \\ \mathbf{B}_{k}^{H} & \mathbf{I}_{N_{\rm s}}\end{array}\right] \succeq \mathbf{0} \text{ and } 
	\left[\begin{array}{cc}\mathbf{Z}_{k} & \mathbf{A}_{k} \\ \mathbf{A}_{k}^{H} & \mathbf{I}_{N_{\rm A}}\end{array}\right] \succeq \mathbf{0}, \, \forall k.
\end{equation} 
Consequently, $\mathcal{P} 4$ can be relaxed into a semi-definite programming (SDP) problem as follows
\begin{align*}
	\mathcal{P} 5: &\min_{\{\mathbf{B}_k,\mathbf{A}_k,\mathbf{X}_k,\mathbf{Z}_k\}_{k=1}^N}  \sum_{k=1}^{N} \Big( \operatorname{Tr} \left( \mathbf{W}_k\mathbf{C}_k^H\bar{\mathbf{H}}_k \mathbf{X}_k \bar{\mathbf{H}}_k^H\mathbf{C}_k\right) \\
	&{}+ \operatorname{Tr} \left( \mathbf{W}_k\mathbf{C}_k^H\bar{\mathbf{H}}_k \mathbf{Z}_k \bar{\mathbf{H}}_k^H\mathbf{C}_k\right) - 2\Re\left\{ \operatorname{Tr} \left( \mathbf{W}_k\mathbf{C}_k^H\bar{\mathbf{H}}_k\mathbf{B}_k\right) \right\}  \\
	& {}+ \operatorname{Tr} \left( \mathbf{W}_k^{\prime}\mathbf{C}_k^{\prime H}\bar{\mathbf{H}}_k^{\prime}\mathbf{Z}_k\bar{\mathbf{H}}_k^{\prime H}\mathbf{C}_k^{\prime}\right) - 2\Re\left\{\operatorname{Tr} \left( \mathbf{W}_k^{\prime}\mathbf{C}_k^{\prime H}\bar{\mathbf{H}}_k^{\prime}\mathbf{A}_k\right)\right\}  \\
	& {}+ \operatorname{Tr} \left( \mathbf{W}_k^{\prime \prime}\bar{\mathbf{H}}_k^{\prime}\mathbf{X}_k \bar{\mathbf{H}}_k^{\prime H}\right) + \operatorname{Tr} \left( \mathbf{W}_k^{\prime \prime}\bar{\mathbf{H}}_k^{\prime} \mathbf{Z}_k\bar{\mathbf{H}}_k^{\prime H}\right)  \Big) \\
	\mbox{s.t.} \\ C_6: &\sum_{k=1}^{N} \operatorname{Tr}\left(\mathbf{X}_k + \mathbf{Z}_k \right)  \leq P_{\rm A},\\
	C_7 : &\sum_{k = 1}^N\operatorname{Tr}\left( \boldsymbol{\Phi}\mathbf{T}_k\left( \mathbf{X}_k + \mathbf{Z}_k\right)\mathbf{T}_k^H\boldsymbol{\Phi}^H \right) + \sigma_{\rm I}^2 \operatorname{Tr}\left( \boldsymbol{\Phi}\boldsymbol{\Phi}^H \right) \leq P_{\rm I},\\
	C_8: &\sum_{k = 1}^{N} \operatorname{Tr}\left( \left(\mathbf{I} - \mathbf{D}\right) \left( \bar{\mathbf{H}}_k\left(\mathbf{X}_k + \mathbf{Z}_k\right)\bar{\mathbf{H}}_k^H \right)\right) \geq \bar{\gamma}\left(E_{\rm th}\right) - \kappa_2,\\
	C_{9}: &\left[\begin{array}{cc}\mathbf{X}_{k} & \mathbf{B}_{k} \\ \mathbf{B}_{k}^{H} & \mathbf{I}_{N_{\rm s}}\end{array}\right] \succeq \mathbf{0}, \,\, \forall k, \\
	C_{10}: & \left[\begin{array}{cc}\mathbf{Z}_{k} & \mathbf{A}_{k} \\ \mathbf{A}_{k}^{H} & \mathbf{I}_{N_{\rm A}}\end{array}\right] \succeq \mathbf{0}, \,\, \forall k,
	\end{align*}
where $\kappa_2 \triangleq \sum_{k = 1}^{N} \operatorname{Tr}\left( \left(\mathbf{I} - \mathbf{D}\right) \left(  \sigma_{\rm I}^2  \mathbf{R}_k\mathbf{\Phi}\mathbf{\Phi}^H\mathbf{R}_k^H + \sigma_{\rm ant}^2\mathbf{I}_{N_{\rm B}}\right)\right)$.
Now, the problem at hand is a convex SDP problem and thus can be solved by using the interior-point method or calling existing convex solvers, e.g., CVX \cite{R4}. Note that the optimal objective value of the relaxed problem $\mathcal{P} 5$ serves as a lower bound of that of the original problem $\mathcal{P} 4$. If the solution to $\mathcal{P} 5$ satisfies $\mathbf{X}_{k}^{\star}=\mathbf{B}_{k}^{\star} (\mathbf{B}_{k}^{\star})^{H}$ and $\mathbf{Z}_{k}^{\star}=\mathbf{A}_{k}^{\star} (\mathbf{A}_{k}^{\star})^{H},\,\forall k$, then the relaxation in $\mathcal{P} 5$ is tight, i.e., $\mathcal{P} 4$ and $\mathcal{P} 5$ are equivalent.

\subsection{Optimization of the PS Matrix $\mathbf{D}$}	
Given blocks $\mathbf{\Xi}$, $\{\mathbf{B}_k,\mathbf{A}_k\}_{k=1}^N$, and $\boldsymbol{\Phi}$, the optimization problem corresponding to the PS matrix $\mathbf{D}$ is given by
\begin{align*}
	\mathcal{P} 6: \, \, \min_{\mathbf{D}} \quad &  \sum_{k=1}^{N}  \operatorname{Tr} \left( \mathbf{W}_k \mathbf{E}_k\right) \\
	\mbox{s.t.} \quad  &C_4, \, \, \, C_5.
\end{align*}
By virtue of the diagonal structure of $\mathbf{D} = \operatorname{diag}(\{d_r\}_{r=1}^{N_{\rm B}})$, it can be further rewritten as
\begin{align*}
	\mathcal{P} 7: \, \, \min_{\{d_r\}_{r=1}^{N_{\rm B}}} \quad & \sum_{r=1}^{N_{\rm B}} \frac{\sigma_{\rm sp}^2 t_{r}}{d_r} \\
	\mbox{s.t.} \,\, \quad &C_{11}: \sum_{r=1}^{N_{\rm B}} d_r \hat{t}_{r} \leq   \sum_{r=1}^{N_{\rm B}} \hat{t}_{r}  - \bar{\gamma}\left(E_{\rm th}\right), \\
	& C_{12}:  0 \leq d_r \leq 1,\,\, \forall r
\end{align*}
where $t_{r}$ and $\hat{t}_{r}$ are the $r$-th diagonal elements of $\sum_{k=1}^{N}\mathbf{C}_k\mathbf{W}_k \mathbf{C}_k^H$ and $\sum_{k=1}^{N}\bar{\mathbf{H}}_k(\mathbf{B}_k\mathbf{B}_k^H + \mathbf{A}_k\mathbf{A}_k^H)\bar{\mathbf{H}}_k^H +  \sigma_{\rm I}^2  \mathbf{R}_k\mathbf{\Phi}\mathbf{\Phi}^H\mathbf{R}_k^H + \sigma_{\rm ant}^2\mathbf{I}_{N_{\rm B}}$, respectively. It can be seen that $\mathcal{P} 7$ is a convex optimization problem with convex objective function and affine constraints. To facilitate a more efficient algorithm design, we introduce a non-negative penalty factor $u$ to make the EH constraint $C_{11}$ implicit in the objective, yielding
\begin{align*}
	 \mathcal{P} 8: \min_{\{d_r\}_{r=1}^{N_{\rm B}}} &\sum_{r=1}^{N_{\rm B}} \frac{\sigma_{\rm sp}^2 t_{r}}{d_r} + u  \left( \sum_{r=1}^{N_{\rm B}} d_r \hat{t}_{r} - \left(\sum_{r=1}^{N_{\rm B}} \hat{t}_{r}  - \bar{\gamma}\left(E_{\rm th}\right)\right) \right) \\
	\quad \operatorname{s.t.} \,\, & C_{12}.
\end{align*}
For a given $u$, $\mathcal{P} 8$ is a convex optimization problem. By using the first-order optimality condition \cite[Ch.~5]{R4}, the optimal solution to $\mathcal{P} 8$ can be derived in a closed-form as
\begin{align}
	\begin{split}
		d_r^{\star}(u) = \left\{
		\begin{array}{ll}
			1,  & \operatorname{if} \,\, u = 0 ;\\
			\min \left\{1, \, \sqrt{\frac{\sigma_{\mathrm{sp}}^2 t_r}{u\hat{t}_r}}\right\},  & \operatorname{if} \,\, u > 0. \label{dopt}
		\end{array}
		\right.
	\end{split}
\end{align}
Next, we need to search for the optimal penalty factor $u$ such that the complementary slackness condition for the EH constraint $C_{11}$ is satisfied \cite[Ch.~5]{R4}, i.e.,
\begin{align}
	u\left( \left(\sum_{r=1}^{N_{\rm B}} \hat{t}_{r}  - \bar{\gamma}\left(E_{\rm th}\right)\right) -  h(u) \right) = 0, \label{slc}
\end{align}
where $h(u) \triangleq \sum_{r=1}^{N_{\rm r}}  d_r^{\star}(u) \hat{t}_{r}$. According to (\ref{dopt}), if $u = 0$, then we have $\mathbf{D} = \mathbf{I}_{N_{\rm B}}$, which means that all the received signal power is used for ID. Therefore, for the considered SWIPT system with EH requirement, $u > 0$ always holds, and thus, the equation (\ref{slc}) holds only if $\left(\sum_{r=1}^{N_{\rm B}} \hat{t}_{r}  - \bar{\gamma}\left(E_{\rm th}\right)\right) - h(u) = 0$. Since $h(u)$ is a monotonically decreasing function of $u$, the bisection method can be employed to find the optimal $u$ efficiently \cite[P.~164]{R4}. To sum up, the overall algorithm for optimizing the PS matrix is formalized in Algorithm \ref{BiSearch}.

\begin{algorithm}[t] \label{BiSearch}
	\small
	\caption{Bisection method for the PS matrix $\mathbf{D}$}
	\LinesNumbered
	\KwIn{$\{\mathbf{H}_k,\mathbf{R}_k,\mathbf{T}_k,\mathbf{C}_k,\mathbf{W}_k,\mathbf{A}_k,\mathbf{B}_k\}_{k=1}^N$, $\boldsymbol{\Phi}$, $\sigma_{ \rm I}^2$, $\sigma_{ \rm ant}^2$, $\sigma_{ \rm sp}^2$, $E_{\rm th}$;}
	
	Initialize $u_l$ and $u_u$;
	
	\Repeat{\rm $(u_u - u_l) < \varepsilon$, where  $\varepsilon > 0$ is the convergence accuracy requirement}
	{
		Set $u_m = (u_l + u_u)/2$;
		
		Obtain $\{d_r^{\star}(u_m)\}_{r=1}^{N_B}$ via (\ref{dopt});
		
		Calculate $h(u_m) =  \sum_{r=1}^{N_{\rm r}}  d_r^{\star}(u_m) \hat{t}_{r}$;
		
		\eIf{$\left(\sum_{r=1}^{N_{\rm B}} \hat{t}_{r}  - \bar{\gamma}\left(E_{\rm th}\right)\right) > h(u_m)$}
		{Set $u_l = u_m$;}
		{Set $u_u = u_m$;}
	}	
	\KwOut{$\mathbf{D} \triangleq \operatorname{diag}(\{d_r^{\star}(u_m)\}_{r=1}^{N_B})$.}
\end{algorithm}

\subsection{Optimization of the IRS Matrix $\boldsymbol{\Phi}$}	
Given blocks $\mathbf{\Xi}$, $\{\mathbf{B}_k,\mathbf{A}_k\}_{k=1}^N$ and $\mathbf{D}$, the optimization problem corresponding to the IRS matrix $\boldsymbol{\Phi}$ is written as
\begin{align*}
	\mathcal{P} 9: \, \, \min_{\mathbf{\Phi}} \quad & \sum_{k=1}^{N} \Big( \operatorname{Tr} \left( \mathbf{W}_k \mathbf{E}_k\right) + \operatorname{Tr} \left( \mathbf{W}_k^{\prime} \mathbf{E}_k^{\prime}\right)\\
	& \quad \quad \,\, + \operatorname{Tr} \left( \bar{\mathbf{W}}_k^{\prime}\bar{\mathbf{E}}_k^{\prime}\right)+ \operatorname{Tr}\left( \mathbf{W}_k^{\prime \prime}\mathbf{E}_k^{\prime \prime}\right)\Big) \\
	\mbox{s.t.} \quad & C_2, \, \, C_5.
\end{align*}
By denoting $\mathbf{K}_{1,k} \triangleq \mathbf{C}_k\mathbf{W}_k\mathbf{C}_k^H$, $\mathbf{K}_{2,k} \triangleq \mathbf{B}_k\mathbf{B}_k^H + \mathbf{A}_k\mathbf{A}_k^H$, $\mathbf{K}_{3,k} \triangleq \mathbf{C}_k\mathbf{W}_k\mathbf{B}_k^H$, $\mathbf{K}_{4,k} \triangleq \mathbf{C}_k^{\prime}\mathbf{W}_k^{\prime}\mathbf{C}_k^{\prime H}$, $\mathbf{K}_{5,k} \triangleq \mathbf{C}_k^{\prime}\mathbf{W}_k^{\prime}\mathbf{A}_k^{H}$ and upon rearrangements, $\mathcal{P} 9$ can be further expanded as $\mathcal{P} 10$, shown at the top of the page.
\begin{figure*}[!t]
	\normalsize
	\setlength{\arraycolsep}{0.0em}
	{ \small \begin{align*}
			\mathcal{P} 10: \, \, \min_{\mathbf{\Phi}} \quad & \sum_{k=1}^{N} \Big( \operatorname{Tr} \left(\mathbf{R}_k^H \mathbf{K}_{1,k} \mathbf{H}_k \mathbf{K}_{2,k} \mathbf{T}_k^H \mathbf{\Phi}^{H}\right) + \operatorname{Tr} \left(\mathbf{T}_k \mathbf{K}_{2,k} \mathbf{H}_k^H \mathbf{K}_{1,k} \mathbf{R}_k \mathbf{\Phi}\right)  + \operatorname{Tr} \left(\mathbf{R}_k^H \mathbf{K}_{1,k} \mathbf{R}_k \mathbf{\Phi} \mathbf{T}_k \mathbf{K}_{2,k}  \mathbf{T}_k^H \mathbf{\Phi}^{H}  \right) \\
			& \quad \quad  + \sigma_{\rm I}^2 \operatorname{Tr} \left(\mathbf{R}_k^H \mathbf{K}_{1,k}\mathbf{R}_k\boldsymbol{\Phi}\boldsymbol{\Phi}^H\right)   - \operatorname{Tr} \left(\mathbf{R}_k^H \mathbf{K}_{3,k}\mathbf{T}_k^H \mathbf{\Phi}^{H}\right)   -  \operatorname{Tr} \left(\mathbf{T}_k \mathbf{K}_{3,k}^H \mathbf{R}_k \mathbf{\Phi}\right) + \operatorname{Tr} \left(\mathbf{R}_k^{\prime H} \mathbf{K}_{4,k} \mathbf{H}_k^{\prime} \mathbf{A}_{k} \mathbf{A}_{k}^H \mathbf{T}_k^H \mathbf{\Phi}^{H}\right) \\
			& \quad \quad + \operatorname{Tr} \left(\mathbf{T}_k \mathbf{A}_{k} \mathbf{A}_{k}^H \mathbf{H}_k^{\prime H} \mathbf{K}_{4,k} \mathbf{R}_k^{\prime} \mathbf{\Phi} \right)  + \operatorname{Tr} \left(\mathbf{R}_k^{\prime H} \mathbf{K}_{4,k} \mathbf{R}_k^{\prime} \mathbf{\Phi} \mathbf{T}_k \mathbf{A}_{k} \mathbf{A}_{k}^H \mathbf{T}_k^H \mathbf{\Phi}^{H}  \right) + \sigma_{\rm I}^2 \operatorname{Tr} \left(\mathbf{R}_k^{\prime H} \mathbf{K}_{4,k}\mathbf{R}_k^{\prime}\boldsymbol{\Phi}\boldsymbol{\Phi}^H\right) \\
			& \quad \quad - \operatorname{Tr} \left(\mathbf{R}_k^{\prime H} \mathbf{K}_{5,k}\mathbf{T}_k^H \mathbf{\Phi}^{H}\right) -  \operatorname{Tr} \left(\mathbf{T}_k \mathbf{K}_{5,k}^H \mathbf{R}_k^{\prime} \mathbf{\Phi} \right) + \operatorname{Tr} \left(\mathbf{R}_k^{\prime H} \bar{\mathbf{C}}_k^{\prime}\bar{\mathbf{W}}_k^{\prime}\bar{\mathbf{C}}_k^{\prime H}\mathbf{R}_k^{\prime}\boldsymbol{\Phi}\boldsymbol{\Phi}^H\right)  - \operatorname{Tr} \left( \mathbf{R}_k^{\prime H}\bar{\mathbf{C}}_k^{\prime}\bar{\mathbf{W}}_k^{\prime}\boldsymbol{\Phi}^H\right) \\
			& \quad \quad - \operatorname{Tr} \left( \bar{\mathbf{W}}_k^{\prime}\bar{\mathbf{C}}_k^{\prime H}\mathbf{R}_k^{\prime}\boldsymbol{\Phi}\right)  + \operatorname{Tr} \left(\mathbf{R}_k^{\prime H} \mathbf{W}_{k}^{\prime \prime} \mathbf{H}_k^{\prime} \mathbf{K}_{2,k} \mathbf{T}_k^H \mathbf{\Phi}^{H} \right)  + \operatorname{Tr} \left(\mathbf{T}_k \mathbf{K}_{2,k} \mathbf{H}_k^{\prime H} \mathbf{W}_{k}^{\prime \prime} \mathbf{R}_k^{\prime} \mathbf{\Phi} \right) \\
			& \quad \quad + \operatorname{Tr} \left(\mathbf{R}_k^{\prime H}\mathbf{W}_{k}^{\prime \prime} \mathbf{R}_k^{\prime} \mathbf{\Phi} \mathbf{T}_k \mathbf{K}_{2,k} \mathbf{T}_k^H \mathbf{\Phi}^{H} \right)  + \sigma_{\rm I}^2  \operatorname{Tr}\left(\mathbf{R}_k^{\prime H} \mathbf{W}_k^{\prime \prime}\mathbf{R}_k^{\prime}\mathbf{\Phi}\mathbf{\Phi}^H\right)\Big) \\
			\mbox{s.t.} \quad &C_2, \,\, C_5.
	\end{align*}}
	\setlength{\arraycolsep}{5pt}
	\hrulefill
\end{figure*}
By further defining $\mathbf{K}_{6,k} \triangleq \mathbf{R}_k^H \mathbf{K}_{1,k} \mathbf{H}_k \mathbf{K}_{2,k} \mathbf{T}_k^H + \mathbf{R}_k^{\prime H}\mathbf{K}_{4,k} \mathbf{H}_k^{\prime} \mathbf{A}_{k} \mathbf{A}_{k}^H \mathbf{T}_k^H + \mathbf{R}_k^{\prime H} \mathbf{W}_{k}^{\prime \prime} \mathbf{H}_k^{\prime} \mathbf{K}_{2,k} \mathbf{T}_k^H - \mathbf{R}_k^H\mathbf{K}_{3,k}\mathbf{T}_k^H - \mathbf{R}_k^{\prime H}\mathbf{K}_{5,k}\mathbf{T}_k^H - \mathbf{R}_k^{\prime H}\bar{\mathbf{C}}_k^{\prime}\bar{\mathbf{W}}_k^{\prime}$, $\mathbf{K}_{7,k} \triangleq \mathbf{R}_k^H \mathbf{K}_{1,k} \mathbf{R}_k  + \sigma_{\rm I}^2 \mathbf{R}_k^H \mathbf{K}_{1,k}\mathbf{R}_k + \mathbf{R}_k^{\prime H} \mathbf{K}_{4,k} \mathbf{R}_k^{\prime}  + \sigma_{\rm I}^2 \mathbf{R}_k^{\prime H} \mathbf{K}_{4,k}\mathbf{R}_k^{\prime} +  \mathbf{R}_k^{\prime H} \bar{\mathbf{C}}_k^{\prime}\bar{\mathbf{W}}_k^{\prime}\bar{\mathbf{C}}_k^{\prime H}\mathbf{R}_k^{\prime} + \mathbf{R}_k^{\prime H} \mathbf{W}_{k}^{\prime \prime} \mathbf{R}_k^{\prime}  + \sigma_{\rm I}^2 \mathbf{R}_k^{\prime H} \mathbf{W}_k^{\prime \prime}\mathbf{R}_k^{\prime}$, and $\mathbf{K}_{8,k} \triangleq \mathbf{T}_k \mathbf{K}_{2,k} \mathbf{T}_k^H + \mathbf{T}_k \mathbf{A}_{k} \mathbf{A}_{k}^H \mathbf{T}_k^H  + \mathbf{T}_k \mathbf{K}_{2,k}\mathbf{T}_k^H + 4\mathbf{I}_{M}$, the objective function of $\mathcal{P} 10$ can be rewritten in a compact form as
\begin{align}
	M(\boldsymbol{\Phi}) & =  \sum_{k = 1}^M \left(  \operatorname{Tr} \left( \mathbf{K}_{7,k}  \mathbf{\Phi} \mathbf{K}_{8,k} \mathbf{\Phi}^{H} \right) + 2 \Re\left\{ \operatorname{Tr} \left( \mathbf{K}_{6,k} \mathbf{\Phi}^{H} \right)  \right\} \right)  \notag \\
	& = \boldsymbol{\phi}^H \mathbf{K} \boldsymbol{\phi} + 2 \Re\left\{ \boldsymbol{\phi}^H \mathbf{q} \right\} , \label{39}
\end{align}
where $\boldsymbol{\phi} = \operatorname{diag}\left( \mathbf{\Phi} \right)$, $\mathbf{K} \triangleq  \sum_{k = 1}^M \mathbf{K}_{7,k} \odot \mathbf{K}_{8,k}^{T}$, and $\mathbf{q} = \sum_{k = 1}^M \operatorname{diag}\left(\mathbf{K}_{6,k}\right)$. Similarly, the constraints $C_2$ and $C_5$ in $\mathcal{P} 10$ can be rewritten in compact forms as
\begin{align}
	&C_{13}: \boldsymbol{\phi}^H \hat{\mathbf{K}} \boldsymbol{\phi}  + \sigma_{\rm I}^2\boldsymbol{\phi}^H \boldsymbol{\phi} \leq P_{\rm I}, \label{40} \\ 
	&C_{14}: \boldsymbol{\phi}^H \bar{\mathbf{K}} \boldsymbol{\phi} + 2 \Re\left\{ \boldsymbol{\phi}^H \bar{\mathbf{q}} \right\} + \kappa_3 \geq  \bar{\gamma}\left(E_{\rm th}\right), \label{41}
\end{align}
where $\hat{\mathbf{K}} \triangleq \sum_{k=1}^{N} \mathbf{I}_M \odot \big(\mathbf{T}_k \mathbf{K}_{2,k} \mathbf{T}_k^H \big)^T$, $\bar{\mathbf{K}} \triangleq \sum_{k=1}^{N} \big(\mathbf{R}_k^H \big(\mathbf{I}_{N_{\rm B}} - \mathbf{D} \big) \mathbf{R}_k \odot \left(\mathbf{T}_k \mathbf{K}_{2,k} \mathbf{T}_k^H\right)^T + \sigma_{\rm I}^2\mathbf{R}_k^H \big(\mathbf{I}_{N_{\rm B}} - \mathbf{D} \big) \mathbf{R}_k \odot \mathbf{I}_{M}\big)$, $\bar{\mathbf{q}} \triangleq \operatorname{diag}\big(\sum_{k=1}^{N}\big(\mathbf{R}_k^H (\mathbf{I}_{N_{\rm B}} - \mathbf{D} ) \mathbf{H}_k \mathbf{K}_{2,k}\mathbf{T}_k^H\big)\big)$, and  $\kappa_3 \triangleq \sum_{r=1}^{N_{\rm B}} (1-d_r)\sigma_{\rm ant}^2N + \sum_{k=1}^{N} \operatorname{Tr} \big( \big(\mathbf{I}_{N_{\rm B}} - \mathbf{D}\big)\mathbf{H}_k \mathbf{K}_{2,k} \mathbf{H}_k^H \big)$. As a result, $\mathcal{P} 10$ can be equivalently rewritten as 
\begin{align*}
	\mathcal{P} 11: \, \, \min_{\boldsymbol{\phi}} \quad  & \boldsymbol{\phi}^H \mathbf{K} \boldsymbol{\phi} + 2 \Re\left\{ \boldsymbol{\phi}^H \mathbf{q} \right\} \\	\mbox{s.t.} \quad &C_{13}, \,\, C_{14}.
\end{align*}
It can be seen that $\mathcal{P} 11$ is a non-convex QCQP problem since the EH constraint $C_{14}$ is a reverse convex constraint. 

In the following, we solve $\mathcal{P} 11$ based on the IA method, and the basic idea is to handle the original problem by solving a sequence of approximating convex problems \cite{R5}. Specifically, since $\boldsymbol{\phi}^H \bar{\mathbf{K}} \boldsymbol{\phi}$ is convex in terms of $\boldsymbol{\phi}$, the first-order Taylor expansion at any feasible point $\tilde{\boldsymbol{\phi}}$ is its lower bound, i.e.,
\begin{align}
	\label{lb1}
	\boldsymbol{\phi}^H \bar{\mathbf{K}} \boldsymbol{\phi} \geq - \tilde{\boldsymbol{\phi}}^H \bar{\mathbf{K}} \tilde{\boldsymbol{\phi}}  +2 \Re \left\{\boldsymbol{\phi}^H \bar{\mathbf{K}} \tilde{\boldsymbol{\phi}} \right\}.
\end{align}
By applying this lower bound to the EH constraint $C_{14}$, we obtain an approximate convex problem as follows:
\begin{align*}
	&\mathcal{P} 12: \, \, \min_{\boldsymbol{\phi}}  \,\, \boldsymbol{\phi}^H \mathbf{K} \boldsymbol{\phi} + 2 \Re\left\{ \boldsymbol{\phi}^H \mathbf{q} \right\} \\
	& \, \, \mbox{s.t.}  \,\, C_{13}, \,\, C_{15}: \, 2 \Re\left\{ \boldsymbol{\phi}^H \left(\bar{\mathbf{K}} \boldsymbol{\phi}^{[n]} + \bar{\mathbf{q}}\right) \right\} \geq  \bar{\gamma}\left(E_{\rm th}\right) - \kappa_4,
\end{align*}
where $\boldsymbol{\phi}^{[n]}$ is the reflection-coefficients vector obtained at the previous iteration, and $\kappa_4 \triangleq \kappa_3 - (\boldsymbol{\phi}^{[n]})^H \bar{\mathbf{K}} \boldsymbol{\phi}^{[n]}$. $\mathcal{P} 12$ is a convex QCQP problem and thus can be effectively solved with the interior-point algorithm or calling existing convex solvers, e.g., CVX \cite{R4}. By successively solving $\mathcal{P} 12$, we can obtain a KKT solution to the original problem $\mathcal{P} 11$ \cite{R5}. In summary, the IA-based algorithm for optimizing the IRS matrix is formalized in Algorithm \ref{IA}.

\begin{algorithm}[t] \label{IA}
	\small
	\caption{IA method for the IRS matrix $\boldsymbol{\Phi}$}
	\LinesNumbered
	\KwIn{$\{\mathbf{H}_k,\mathbf{R}_k,\mathbf{T}_k, \mathbf{H}_k^{\prime}, \mathbf{R}_k^{\prime}\}_{k=1}^N$, $\boldsymbol{\Xi}$, $\sigma_{ \rm I}^2$, $\sigma_{ \rm ant}^2$, $\sigma_{ \rm sp}^2$, $P_{\rm I}$, $E_{\rm th}$;}
	
	Initialize $\boldsymbol{\phi}^{[0]}$ and set $n = 0$;
	
	\Repeat{\rm $(M^{[n-1]} - M^{[n]})/M^{[n-1]} \leq \varepsilon$, where  $\varepsilon > 0$ is the convergence accuracy requirement}
	{
		Set $n = n+1$;
		
		Obtain $\boldsymbol{\phi}^{[n]}$ by solving $\mathcal{P}12$;
		
		Calculate the objective value of $\mathcal{P}11$, denoted as $M^{[n]}$;		
	}	
	\KwOut{$\boldsymbol{\Phi} \triangleq \operatorname{diag}(\boldsymbol{\phi}^{[n]})$.}
\end{algorithm}

\subsection{Algorithm Summary and Complexity Analysis}
To sum up, Algorithm \ref{BCD} formalizes the overall BCD-based algorithm for solving $\mathcal{P}2$ (or equivalently $\mathcal{P}1$). It can be seen that Steps 4-7 yield a non-decreasing objective value of $\mathcal{P}2$ over the iterations, which, meanwhile, are upper bounded by the transmit and reflecting power constraints. Therefore, given a feasible initial point, Algorithm \ref{BCD} is guaranteed to converge. 

\begin{algorithm}[t] \label{BCD}
	\small
	\caption{BCD-based algorithm for $\mathcal{P}2$ (or $\mathcal{P}1$)}
	\LinesNumbered
	\KwIn{$\{\mathbf{H}_k,\mathbf{R}_k,\mathbf{T}_k,\mathbf{H}_k^{\prime},\mathbf{R}_k^{\prime}\}_{k=1}^N$, $\sigma_{ \rm I}^2$, $\sigma_{ \rm ant}^2$, $\sigma_{\rm sp}^2$, $\sigma_{\rm E}^2$, $P_{\rm A}$, $P_{\rm I}$, $E_{\rm th}$;}
	
	Initialize $\{\mathbf{B}_k^{[0]},\mathbf{A}_k^{[0]}\}_{k=1}^N,\mathbf{\Phi}^{[0]},\mathbf{D}^{[0]}$, and set $t=0$;
	
	\Repeat{\rm $(R_{\rm Sec}^{[t]}-R_{\rm Sec}^{[t-1]})/R_{\rm Sec}^{[t]} < \varepsilon$, where  $\varepsilon > 0$ is the convergence accuracy requirement}
	{
		Set $t = t+1$;
		
		Update the auxiliary variables $\mathbf{\Xi}^{[t]}$ via (\ref{a1}) - (\ref{wopt});
		
		Optimize the information/AN precoding matrices $\{\mathbf{B}_k^{[t]},\mathbf{A}_k^{[t]}\}_{k=1}^N$ via solving $\mathcal{P}5$;
		
		Optimize the PS matrix $\mathbf{D}^{[t]}$ via Algorithm \ref{BiSearch};
		
		Optimize the IRS matrix $\boldsymbol{\Phi}^{[t]}$ via Algorithm \ref{IA};
		
		Calculate the objective value of $\mathcal{P}2$ (or $\mathcal{P}1$), denoted as $R_{\rm Sec}^{[t]}$.
	}
	
	\KwOut{$\{\mathbf{B}_k^{[t]},\mathbf{A}_k^{[t]}\}_{k=1}^N,\mathbf{\Phi}^{[t]},\mathbf{D}^{[t]}$;}
\end{algorithm}

\textit{\underline{Complexity Analysis}:} At each iteration, since the optimal $\boldsymbol{\Xi}$ and $\mathbf{D}$ can be derived in closed/semi-closed forms, the major computational complexity lies in optimizing the blocks $\{\mathbf{B}_k,\mathbf{A}_k\}_{k=1}^N$ and $\boldsymbol{\Phi}$, i.e., Step 5 and Step 7 in Algorithm \ref{BCD}. Specifically, for optimizing the block $\{\mathbf{B}_k,\mathbf{A}_k\}_{k=1}^N$, the interior-point algorithm is employed to solve the SDP problem $\mathcal{P}5$, whose computational complexity is given by $\mathcal{O}\left(N N_{\rm A}^{4.5} \log (1 / \epsilon)\right)$, where $\epsilon$ is the solution accuracy. For optimizing the block $\boldsymbol{\Phi}$, we need to iteratively solve the QCQP problem $\mathcal{P}12$, whose computational complexity is given by $\mathcal{O}(N_{\rm IA,1}M^{3} \log (1 / \epsilon))$, where $N_{\rm IA}$ is the iteration number of Algorithm \ref{IA}. Therefore, the overall computational complexity of Algorithm \ref{BCD} is approximately given by $\mathcal{O}(N_{\rm BCD}(N N_{\rm A}^{4.5} \log (1 / \epsilon) + N_{\rm IA,1}M^{3} \log (1 / \epsilon)))$, where $N_{\rm BCD}$ represents the iteration number of Algorithm \ref{BCD}.

\section{Low-Complexity Algorithm}
\label{Sec-lowCompAl}
The BCD-based algorithm proposed in the previous section exhibits relatively high computational complexity since it takes many iterations to converge. We need to solve each iteration's SDP problem with high computational complexity. This section proposes a three-stage optimization strategy to solve $\mathcal{P}1$ with low complexity. First, we optimize the IRS matrix to maximize the channel gain of the legitimate user. Then, a ZF-based scheme is designed to optimize the precoding matrices efficiently. Finally, we adjust the PS matrix to further improve the secrecy sum rate.

In the first stage, we optimize the IRS matrix so that the combined channel gains over all subcarriers of the legitimate user are maximized, which can be formulated as
\begin{align*}
	\mathcal{P} 13: \,\max_{\mathbf{\Phi}} \,  &\sum_{k = 1}^N \| \mathbf{H}_k + \mathbf{R}_k \boldsymbol{\Phi} \mathbf{T}_k \|_F^2  \\
	\mbox{s.t.} \quad C_{16}: \, &\tilde{P}_{\rm A}\sum_{k = 1}^N\operatorname{Tr}\left( \boldsymbol{\Phi}\mathbf{T}_k\mathbf{T}_k^H\boldsymbol{\Phi}^H \right) + \sigma_{\rm I}^2 \operatorname{Tr}\left( \boldsymbol{\Phi}\boldsymbol{\Phi}^H \right) \leq P_{\rm I},
\end{align*}
where $C_{16}$ imposes the active-IRS's reflecting power constraint, in which we assume the equal power allocation scheme is applied at Alice, i.e., $\tilde{P}_{\rm A} \triangleq  P_{\rm A}/(NN_{\rm t})$. Upon rearrangements, $\mathcal{P} 13$ can be rewritten in a compact form as
\begin{align*}
	\mathcal{P} 14: \, \, \, \max_{\boldsymbol{\phi}} \,\,  & \boldsymbol{\phi}^H \mathbf{J} \boldsymbol{\phi} + 2 \Re\left( \boldsymbol{\phi}^H \mathbf{j} \right)  \\
	\mbox{s.t.} \quad  C_{17}: \, &\boldsymbol{\phi}^H \hat{\mathbf{J}} \boldsymbol{\phi}  + \sigma_{\rm I}^2\boldsymbol{\phi}^H \boldsymbol{\phi} \leq P_{\rm I},
\end{align*}
where $\mathbf{J} \triangleq \sum_{k=1}^N ( \mathbf{R}_k^H\mathbf{R}_k ) \odot ( \mathbf{T}_k\mathbf{T}_k^H )^T$, $\mathbf{j} \triangleq \sum_{k=1}^N \operatorname{diag}( \mathbf{R}_k^H \mathbf{H}_k \mathbf{T}_k^H)$, and $\hat{\mathbf{J}} \triangleq \tilde{P_{\rm A}}\sum_{k=1}^{N} \mathbf{I}_M \odot (\mathbf{T}_k\mathbf{T}_k^H )^T$. However, $\mathcal{P} 14$ cannot be solved directly since its objective is to maximize a convex function. Therefore, we resort to the IA method. Specifically, we construct a surrogate objective function by using the lower bound shown in (\ref{lb1}), and then we obtain
\begin{align*}
	\mathcal{P} 15: \,\, \max_{\boldsymbol{\phi}} \,\,  &  2 \Re( \boldsymbol{\phi}^H (\mathbf{j} + \mathbf{J} \boldsymbol{\phi}^{[n-1]}) ) \quad \mbox{s.t.} \,\, C_{17}.
\end{align*}
By iteratively solving $\mathcal{P} 15$, we can obtain a KKT solution to $\mathcal{P} 14$. Applying the IA method for solving $\mathcal{P} 14$ is similar to Algorithm \ref{IA} and thus omitted here for brevity.

In the second stage, we design the precoding matrices with the obtained IRS matrix to avoid information leakage to the eavesdropper. Toward this end, we choose the information precoding matrices $\{\mathbf{B}_k\}_{k=1}^{N}$ from the null space of the eavesdropper's CFR matrices $\{\bar{\mathbf{H}}_{k}^{\prime}\}_{k=1}^{N}$, i.e.,
\begin{align}
	\mathbf{B}_k = \tilde{\mathbf{V}}_{k} \mathbf{Q}_k, \,\, \forall k \label{zfsdfdd}
\end{align}
where $\tilde{\mathbf{V}}_{k}\in\mathbb{C}^{N_{\rm A} \times (N_{\rm A}-N_{\rm E})}$ is the projection matrix obtained by performing SVD to $\bar{\mathbf{H}}_{k}^{\prime}$, i.e., 
\begin{align}
	\bar{\mathbf{H}}_{k}^{\prime} = \mathbf{U}_{k} \left[\mathbf{\Lambda}_{k},\mathbf{0}\right]\left[\begin{array}{c}
		\mathbf{V}_{k}^{H} \\
		\tilde{\mathbf{V}}_{k}^{H}
	\end{array}\right], \label{zfsdfddssss}
\end{align}
and $\mathbf{Q}_k \in \mathbb{C}^{(N_{\rm A}-N_{\rm E})\times N_{\rm S}}$ is the precoding matrix of lower dimension that can be optimized. It is noteworthy that, with the structure (\ref{zfsdfdd}), the eavesdropping data rate at Eve is forced to zero, and thus the optimal AN precoding matrices $\{\mathbf{A}_k\}_{k=1}^N$ should be set as zero matrices. Besides, the number of spatial streams that can be transmitted on each subcarrier at Alice reduces to $N_{\rm S}=\min\{N_{\rm A}, N_{\rm B}, N_{\rm A}-N_{\rm E}\}$.

By applying (\ref{zfsdfdd}) to $\mathcal{P}1$, the optimization problem corresponding to the precoding matrices reduces to
\begin{align*}
	\mathcal{P} 16: \, \, \max_{\{\mathbf{Q}_k\}_{k=1}^{N}} \, \,  &\sum_{k=1}^{N} \operatorname{log}_2 \operatorname{det} \left( \mathbf{I}_{N_{\rm B}} + \mathbf{G}_k\mathbf{Q}_k\mathbf{Q}_k^H\mathbf{G}_k^H\boldsymbol{\Gamma}_k^{-1}\right)\\
	\mbox{s.t.} \quad C_{18}: \, \,  &\sum_{k=1}^{N} \operatorname{Tr}\big(\mathbf{Q}_k \mathbf{Q}_k^H \big)  \leq P_{\rm A},\\
	 \,\, C_{19}: \, &\sum_{k = 1}^N\operatorname{Tr}\left( \boldsymbol{\Phi}\mathbf{T}_k\left( \tilde{\mathbf{V}}_k\mathbf{Q}_k\mathbf{Q}_k^H \tilde{\mathbf{V}}_k^H\right)\mathbf{T}_k^H\boldsymbol{\Phi}^H \right) \\
	& \qquad {}+ \sigma_{\rm I}^2 \operatorname{Tr}\left( \boldsymbol{\Phi}\boldsymbol{\Phi}^H \right) \leq P_{\rm I},\\
	\,\, C_{20}: \,  &\sum_{k = 1}^{N} \operatorname{Tr}\Big( \left(\mathbf{I} - \mathbf{D}\right) \big( \mathbf{G}_k \mathbf{Q}_k\mathbf{Q}_k^H\mathbf{G}_k^H \\
	& \quad {}+  \sigma_{\rm I}^2  \mathbf{R}_k\mathbf{\Phi}\mathbf{\Phi}^H\mathbf{R}_k^H + \sigma_{\rm ant}^2\mathbf{I}_{N_{\rm B}}\big)\Big) \geq \bar{\gamma}\left(E_{\rm th}\right),
\end{align*}
where $\mathbf{G}_k \triangleq \bar{\mathbf{H}}_k\tilde{\mathbf{V}}_k$, $\forall k$. Based on Hadamard’s inequality on matrix determinant \cite{9598828}, the secrecy rate on each subcarrier $k$ can be maximized if $\left( \mathbf{I}_{N_{\rm B}} + \mathbf{G}_k\mathbf{Q}_k\mathbf{Q}_k^H\mathbf{G}_k^H\boldsymbol{\Gamma}_k^{-1}\right)$ is diagonalized. Therefore, to facilitate a low-complexity algorithm design, we design $\{\mathbf{Q}_k\}_{k=1}^{N}$ such that
\begin{align}
	\mathbf{Q}_k^{*} = \bar{\mathbf{U}}_k \mathbf{\Sigma}_k, \,\, \forall k \label{str}
\end{align}
where $\bar{\mathbf{U}}_k$ consists of the eigenvectors of $\mathbf{G}_k^H\boldsymbol{\Gamma}_k^{-1}\mathbf{G}_k$ corresponding to the $N_{\rm S}$ largest eigenvalues, and $\mathbf{\Sigma}_k\triangleq \operatorname{diag}\left(\{\sqrt{z_{k,i}}\}_{i=1}^{N_{\rm S}}\right)$ represents the power allocation scheme on subcarrier $k$. As a result, with (\ref{str}), $\mathcal{P} 16$ reduces to a power allocation optimization problem as follows:
\begin{align*}
	\mathcal{P} 17: \, \, \max_{\{z_{k,i} \geq 0\}} \, \,\, \, & \sum_{k=1}^{N} \sum_{i=1}^{N_{\rm S}}\operatorname{log}_2 \left( 1+\lambda_{k,i}z_{k,i} \right)\\
	\mbox{s.t.} \quad C_{21}: \,  &\sum_{k=1}^{N} \sum_{i=1}^{N_{\rm S}} z_{k,i}  \leq P_{\rm A},\\
	C_{22}: \,  &\sum_{k=1}^{N} \sum_{i=1}^{N_{\rm S}} z_{k,i} \hat{d}_{k,i} \leq P_{\rm I} - \sigma_{\rm I}^2 \operatorname{Tr}\left( \boldsymbol{\Phi}\boldsymbol{\Phi}^H \right),\\
	C_{23}: \, &\sum_{k=1}^{N} \sum_{i=1}^{N_{\rm S}} z_{k,i} \bar{d}_{k,i} \geq \bar{\gamma}\left(E_{\rm th}\right) - \kappa_5,
\end{align*}
where $\lambda_{k,i}$ denotes the $i$-th largest eigenvalue of $\mathbf{G}_k^H\boldsymbol{\Gamma}_k^{-1}\mathbf{G}_k$; $\hat{d}_{k,i}$ and $\bar{d}_{k,i}$ represent the $i$-th diagonal elements of $\mathbf{U}_k^H\tilde{\mathbf{V}}_k^H\mathbf{T}_k^H\boldsymbol{\Phi}^H\boldsymbol{\Phi}\mathbf{T}_k\tilde{\mathbf{V}}_k\mathbf{U}_k$ and $\mathbf{U}_k^H\tilde{\mathbf{V}}_k^H\bar{\mathbf{H}}_k^H(\mathbf{I} - \mathbf{D})\bar{\mathbf{H}}_k\tilde{\mathbf{V}}_k\mathbf{U}_k$, respectively, and $\kappa_5 \triangleq \sum_{k = 1}^{N} \operatorname{Tr}( (\mathbf{I} - \mathbf{D}) (\sigma_{\rm I}^2  \mathbf{R}_k\mathbf{\Phi}\mathbf{\Phi}^H\mathbf{R}_k^H + \sigma_{\rm ant}^2\mathbf{I}_{N_{\rm B}}) )$. Since $\mathcal{P} 17$ is a convex problem with scalar variables, it can be solved efficiently by the Lagrangian duality method \cite{R4}.

In the third stage, we optimize the PS matrix to maximize the secrecy sum rate. Since the ZF-based precoding scheme is applied in the second stage, the eavesdropping data rate at Eve is forced to zero. Thus, the problem $\mathcal{P} 1$ corresponding to the PS matrix reduces to
\begin{align*}
	\mathcal{P} 18: \, \, \max_{\mathbf{D}} \, \,\, \, &  \sum_{k=1}^{N} \operatorname{log}_2 \operatorname{det} \left( \mathbf{I}_{N_{\rm B}} + \bar{\mathbf{H}}_k\mathbf{B}_k\mathbf{B}_k^H\bar{\mathbf{H}}_k^H\mathbf{\Gamma}_k^{-1}\right) \\
	\mbox{s.t.} \quad  &C_{3}, \quad C_{5},
\end{align*}
where $\mathbf{\Gamma}_k \triangleq \sigma_{\rm I}^2  \mathbf{R}_k\mathbf{\Phi}\mathbf{\Phi}^H\mathbf{R}_k^H + \sigma_{\mathrm{ant}}^2 \mathbf{I}_{N_{\rm B}} + \sigma_{\mathrm{sp}}^2 \mathbf{D}^{-1}$. By introducing auxiliary variable $\hat{d}_{r}=\sigma_{\mathrm{sp}}^2 d_{r}^{-1},\forall r$, and $\hat{\mathbf{D}} = \operatorname{diag} (\{\hat{d}_{r}\}_{r=1}^{N_{\rm B}})$, $\mathcal{P} 18$ can be rewritten as
\begin{align*}
	\mathcal{P} 19: \, \, \max_{\hat{\mathbf{D}}} \, \, & \sum_{k=1}^{N} \operatorname{log}_2 \operatorname{det} \left( \mathbf{M}_k +  \hat{\mathbf{D}} \right) - \operatorname{log}_2 \operatorname{det} \left( \hat{\mathbf{M}}_k + \hat{\mathbf{D}} \right) \\
	\mbox{s.t.} \quad \, C_{24}: \,\,  &[\hat{\mathbf{D}}]_{r,r} \geq \sigma_{ \rm sp}^2, \,\,  \forall r\\
	C_{25}: \, \, &\sum_{r = 1}^{N_{\rm B}} \frac{\sigma_{ \rm sp}^2}{\hat{d}_r} [\mathbf{S}]_{r,r}\leq \operatorname{Tr}\left(\mathbf{S}\right) - \bar{\gamma}\left(E_{\rm th}\right),
\end{align*} 
where $\hat{\mathbf{M}}_k \triangleq \sigma_{\rm I}^2 \mathbf{R}_k\mathbf{\Phi}\mathbf{\Phi}^H\mathbf{R}_k^H + \sigma_{\mathrm{ant}}^2 \mathbf{I}_{N_{\rm B}}$, $\mathbf{M}_k \triangleq \mathbf{G}_k\mathbf{Q}_k\mathbf{Q}_k^H\mathbf{G}_k^H + \hat{\mathbf{M}}_k $, and $\mathbf{S}  \triangleq \sum_{k = 1}^{N} \mathbf{M}_k$. It can be seen that the objective function of $\mathcal{P} 18$ is in a difference-of-concave (DC) form. To handle this difficulty, we resort to the IA method. Specifically, since the function $\log \operatorname{det}(\boldsymbol{\Sigma})$ is concave, by using the first-order Taylor expansion, it can be upper-bounded as
\begin{align}
	\log \operatorname{det}(\boldsymbol{\Sigma}) \leq \log \operatorname{det}\left(\boldsymbol{\Sigma}_t\right)+\operatorname{Tr}\left(\boldsymbol{\Sigma}_t^{-1}\left(\boldsymbol{\Sigma}-\boldsymbol{\Sigma}_t\right)\right). \label{ubbb}
\end{align}
By this upper bound, we construct a surrogate objective function as follows:
\begin{align*}
	\mathcal{P} 20: \, \, \max_{\hat{\mathbf{D}}} \, \,\, \, & 	\sum_{k=1}^{N} \operatorname{log}_2 \operatorname{det} \left( \mathbf{M}_k +  \hat{\mathbf{D}} \right) \\
	& \qquad - \operatorname{Tr}\left( \left( \hat{\mathbf{M}}_k + \hat{\mathbf{D}}^{[n-1]} \right)^{-1} \hat{\mathbf{D}}\right)\\
	\mbox{s.t.} \quad &C_{24}, \quad C_{25}.
\end{align*}
This convex problem can be solved efficiently using the Lagrangian method. By iteratively solving $\mathcal{P} 20$, we can obtain a KKT solution to $\mathcal{P} 19$. The procedures of applying the IA method for solving $\mathcal{P} 19$ is similar to Algorithm \ref{IA} and thus omitted here for brevity.

To sum up, the low-complexity algorithm is formalized in Algorithm \ref{Lowcompl}. Note that Algorithm \ref{Lowcompl} is a single-round optimization algorithm. Because of the zero-forcing precoding scheme, the alternating optimization (AO) process between the IRS matrix, precoding matrices, and PS matrix cannot provide additional performance improvement. This is because modifying the IRS/PS matrix during the AO process would change the zero-forcing structures (see Eq.~\eqref{str}). The computational complexity of Algorithm \ref{Lowcompl} is analyzed below. In the first stage, we need to solve the problem $\mathcal{P} 14$ based on the IA method, whose computational complexity is $\mathcal{O}(N_{\rm IA,2}M^{3} \log (1 / \epsilon))$, where $N_{\rm IA,2}$ is the iteration number in the first stage. In the second stage, the dominant computational complexity lies in solving the problem $\mathcal{P} 17$, whose computational complexity is $\mathcal{O}(NN_{\rm S} \log (1 / \epsilon))$. In the third stage, we need to solve the problem $\mathcal{P} 18$ based on the IA method, whose computational complexity is $\mathcal{O}(N_{\rm IA,3}N_{\rm B}^{3} \log (1 / \epsilon))$, where $N_{\rm IA,3}$ is the iteration number in this stage. Therefore, the overall computational complexity of Algorithm \ref{Lowcompl} is $\mathcal{O}(N_{\rm IA,2}M^{3} \log (1 / \epsilon) + NN_{\rm S} \log (1 / \epsilon) + N_{\rm IA,3}N_{\rm B}^{3} \log (1 / \epsilon))$.  

For clarity, the computational complexities of Algorithms \ref{BCD} \& \ref{Lowcompl} are summarized in Table \ref{table_complexity}. Notice that, unlike Algorithm \ref{BCD}, which requires tens of iterations $N_{\rm BCD}$ to converge (c.f. Fig.~\ref{conver} in Section V), Algorithm \ref{Lowcompl} is a one-round optimization strategy. Moreover, in Algorithm \ref{Lowcompl}, the matrix-based precoding optimization problem is transformed into a scalar-based power allocation optimization problem, which incurs lower computational complexity than the SDR method used in Algorithm \ref{BCD}. Therefore, the overall computational complexity of Algorithm \ref{Lowcompl} is significantly lower than that of Algorithm~\ref{BCD}.

\begin{remark}
	The low-complexity algorithm can be readily extended to the multi-eavesdropper case. Specifically, the IRS and PS matrix optimizations are similar to those in the single-eavesdropper case. To optimize the precoding matrices, we need to stack the CFR matrices of all eavesdroppers into a large matrix and then find the null-space projection matrix of this large matrix to satisfy the zero-forcing precoding structure.
\end{remark}

\begin{table*}[t] 
	\small
	\renewcommand{\arraystretch}{1.4}
	\caption{Complexity Comparison}
	\label{table_complexity}
	\centering
	\begin{tabular}{c|c}
		\toprule[1.5pt]
		\textbf{Algorithm}          	&  \textbf{Computational Complexity}	 		\\
		\hline
		BCD-based Algorithm 3          & $\mathcal{O}(N_{\rm BCD}(N N_{\rm A}^{4.5} \log (1 / \epsilon) + N_{\rm IA,1}M^{3} \log (1 / \epsilon)))$  \\
		\hline
		Low-complexity Algorithm 4     &   $\mathcal{O}( NN_{\rm S} \log (1 / \epsilon) + N_{\rm IA,2}M^{3} + N_{\rm IA,3}N_{\rm B}^{3} \log (1 / \epsilon)  \log (1 / \epsilon))$ \\
		\bottomrule[1.5pt]  
	\end{tabular}
\end{table*}

\begin{algorithm}[t] \label{Lowcompl}
	\small
	\caption{Low-complexity algorithm for $\mathcal{P}1$}
	\LinesNumbered
	\KwIn{$\{\mathbf{H}_k,\mathbf{R}_k,\mathbf{T}_k,\mathbf{H}_k^{\prime},\mathbf{R}_k^{\prime}\}_{k=1}^N$, $\sigma_{ \rm I}^2$, $\sigma_{ \rm ant}^2$, $\sigma_{\rm sp}^2$, $\sigma_{\rm E}^2$, $P_{\rm A}$, $P_{\rm I}$, $E_{\rm th}$;}
	
	\textbf{Stage 1: Maximize the channel gain of the legitimate user by adjusting the IRS matrix:}
	
	Solve the problem $\mathcal{P} 15$ iteratively to obtain the optimal $\boldsymbol{\phi}^{*}$;
	
	Output the obtained IRS matrix as $\boldsymbol{\Phi}^{*} = \operatorname{diag}\{\boldsymbol{\phi}^{*}\}$.
	
	\textbf{Stage 2: Enforce the eavesdropping rate at Eve to zero by designing the precoding matrices:}
	
	Perform SVDs to $\{\bar{\mathbf{H}}_{k}^{\prime}\}_{k=1}^{N}$, and then perform EVDs to $\{\mathbf{G}_k^H\boldsymbol{\Gamma}_k^{-1}\mathbf{G}_k\}_{k=1}^{N}$;
	
	Solve the problem $\mathcal{P} 17$, and then substitute the optimal solution back into (\ref{str}) to obtain $\{\mathbf{Q}_k^{*}\}_{k=1}^{N}$;
	
	Construct the precoding matrices $\{\mathbf{B}_k^{*}\}_{k=1}^{N}$ via \eqref{zfsdfdd};
	
	Output the obtained precoding matrices $\{\mathbf{B}_k^{*}\}_{k=1}^{N}$;
	
	\textbf{Stage 3: Improve the secrecy sum-rate by optimizing the PS matrix:}
	
	Solve the problem $\mathcal{P} 20$ iteratively to obtain the optimal $\hat{\mathbf{D}}^{*}$;
	
	Output the obtained PS matrix as $\mathbf{D}^{*} = \sigma_{\rm sp}^2(\hat{\mathbf{D}}^{*})^{-1}$.
	
\end{algorithm}

\section{Simulation Results and Discussions}
\label{Simul}
\begin{figure}[t] 
	\centering
	\includegraphics[width=0.45\textwidth]{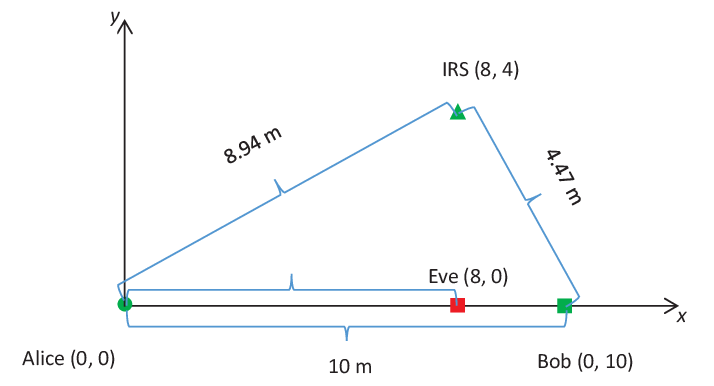}
	\caption{The geometric setting of users and IRS in the simulation experiments.}
	\label{HorLoc}
\end{figure}

In this section, we numerically evaluate the performance of the proposed algorithms compared to benchmark schemes. In the simulation experiments, as shown in Fig. \ref{HorLoc}, Alice, IRS, Bob, and Eve are located in 2-dimensional space at $(0,0) \, \rm{m}$, $(8,4) \, \rm{m}$, $(10,0) \, \rm{m}$, and $(8,0) \, \rm{m}$, respectively. We consider a narrow-band internet-of-thing (NB-IoT) downlink scenario in our simulations \cite{36211}. Unless specified otherwise, the simulation parameters are given in Table \ref{table_simulParam}. For the large-scale fading, the distance-dependent path-loss is modeled as \cite{10280714}: $\mathcal{G}(d) = (\lambda_{\rm c}/4 \pi)^2 (d/D_0)^{-\eta}$, where $d$ denotes the transmission distance, $D_0 = 1 \, \mathrm{m}$ denotes the reference distance, and $\eta$ denotes the path loss exponent. Unless specified otherwise, the path-loss exponents of the Alice-Bob, Alice-IRS, IRS-Bob, Alice-Eve, and IRS-Eve links are assumed to be $\eta_{\rm AB} = 3.2$, $\eta_{\rm AI} = \eta_{\rm IB} = 2.2$, $\eta_{\rm AE} = 3.2$, and $\eta_{\rm IE} = 2.8$, respectively. For the small-scale fading, we utilize the Extended Typical Urban (ETU) model from the 3GPP standard to characterize the delay power profile of multi-path channels \cite{36104}.

\begin{table}[t] \small
	\renewcommand{\arraystretch}{1} \scriptsize
	\caption{Simulation Parameters}
	\label{table_simulParam}
	\centering
	\begin{tabular}{c||c}
		\toprule[1.5pt]
		\textbf{Parameters}             									& \textbf{Values} 		\\
		\hline
		Carrier frequency, $f_{\rm c}$   									& $750 \, \mathrm{MHz}$	\\
		
		Available bandwidth              &    $180 \, \mathrm{kHz}$	\\
		
		Subcarrier spacing      &   $15 \, \mathrm{kHz}$	\\
		
		\# of subcarriers, $N$ 											& $12$					\\
		
		FFT and IFFT points      & 2048 \\
		
		Sampling rate           & $30.72 \, \mathrm{MHz}$	\\
		
		\# of Tx antennas at Alice, $N_{\rm A}$  					& $4$					\\
		
		\# of Rx antennas at Bob, $N_{\rm B}$						& $2$ 					\\
		
		\# of Rx antennas at Eve, $N_{\rm E}$						& $2$ 					\\
		
		\# of reflecting elements, $M$ 									& $40$ 					\\
	
		Noise power at the active IRS, $\sigma_{ \rm I}^2$					& $-40 \, \mathrm{dBm}$  \\
		
		Noise power at Bob, $(\sigma_{ \rm ant}^2, \sigma_{ \rm sp}^2)$    & $(-60, -40) \, \mathrm{dBm}$ \\
		
		Noise power at Eve, $\sigma_{ \rm E}^2$        & $-60 \, \mathrm{dBm}$\\
		
		Non-linear EH model parameters, $(\xi, \nu)$                         & $(274,0.29)$			\\
		
		Saturation power level of the EH circuits, $E_{\rm m}$              & $4.927 \, \mathrm{mW}$ \\
		
		Activation power level of the EH circuits, $E_{\rm 0}$              & $0.064 \, \mathrm{mW}$ \\
		
	    Power delay profile of channels & ETU model \cite{36104}           \\
		\bottomrule[1.5pt]  
	\end{tabular}
\end{table}

For comparison, we include the following benchmark schemes:
\begin{itemize}
	\item \textbf{Passive-IRS, BCD-based alg.:} The problem formulation for the passive-IRS case can be found in Remark \ref{rmk3}. Within the BCD framework, the optimizations for the auxiliary variables, precoding matrices, and PS matrix are similar to those in the active-IRS case. For optimizing the passive-IRS matrix, we first linearize the EH constraint and the objective function by employing the lower-bound \eqref{lb1} and the upper-bound in \cite[~Lemma 1]{8741198}, respectively. Then, we add the EH constraint to the objective by using the penalty method, based on which the solution to the IRS matrix can be obtained in a semi-closed form. Finally, the bisection search method can be applied as in Algorithm 1 to find the optimal penalty factor.
	\item \textbf{Passive-IRS, low-complexity alg.:} In the first stage, the reflecting power constraint $C_{17}$ in $\mathcal{P}15$ is replaced by the uni-modulus constraints (c.f. Remark \ref{rmk3}), and thus the optimal IRS vector can be obtained in a closed-form as $\boldsymbol{\phi} = \exp\left(j\arg\left(\mathbf{j} + \mathbf{J} \boldsymbol{\phi}^{[n-1]}\right)\right)$. In the second and third stages, the precoding matrices and the PS matrix optimizations are similar to those in the active-IRS case.
	\item \textbf{Non-IRS, BCD-based alg.:} With the IRS matrix fixed as a zero matrix, the BCD-based algorithm is the same as for the passive-IRS case.
	\item \textbf{Non-IRS, low-complexity alg.:} With the IRS matrix fixed as a zero matrix, the low-complexity algorithm is the same as for the passive-IRS case.
	\item \textbf{Active-/passive-/non-IRS, Rate maximization (RM):} Assuming that $\{\bar{\mathbf{H}}_k^{\prime} = \mathbf{0}\}_{k=1}^{N}$, i.e., ignoring Eve, and maximizing the sum-rate of the legitimate receiver by using the BCD-based algorithm.
\end{itemize}
Note that, compared to the passive- and non-IRS schemes, an extra power budget, i.e., $P_{\rm I}$, is provided at the reflecting array for the active IRS scheme. Therefore, to make fair comparisons, the transmitter Alice is compensated with an additional transmit power budget, i.e., $(P_{\rm A}+P_{\rm I})$, for the passive- and non-IRS schemes \cite{9849458}.  

\begin{figure}[t]
	\centering
	\includegraphics[width=0.45\textwidth]{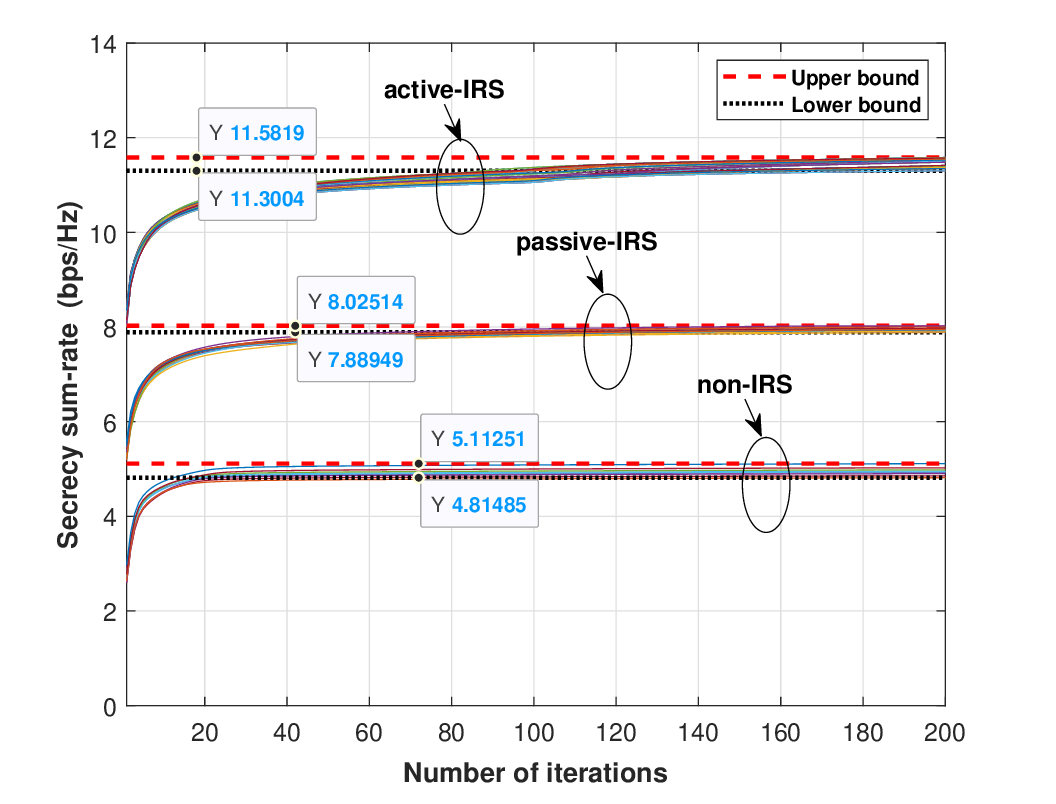}
	\caption{Convergence behavior of the BCD-based algorithms with random initial blocks.}
	\label{conver}
\end{figure}

To start with, Fig.~\ref{conver} illustrates the convergence behavior of the proposed BCD-based algorithm developed in Section~\ref{Sec-BCD}. Specifically, we show the secrecy sum rate versus the number of iterations for the active-, passive-, and non-IRS schemes, with $P_{\rm A} = 34 \,\text{dBm}$, $P_{\rm I} = 22 \,\text{dBm}$, and $E_{\rm th} = -10 \,\text{dBm}$. To gain insights into the optimality gap of the BCD framework, we randomly generate $30$ initial blocks for each scheme, allowing the BCD framework to converge to different locally optimal solutions. It is observed from Fig.~\ref{conver} that the secrecy sum rate achieved by each scheme increases with iterations and becomes saturated when the number of iterations becomes large. This observation indicates that the proposed BCD-based solution is convergent. Additionally, it can be easily computed that the relative optimality gap of the BCD framework for the active-, passive-, and non-IRS schemes are about $2.5\%$, $1.7\%$, and $6.2\%$, respectively.

\begin{figure}[t]
	\centering
	\includegraphics[scale=0.45]{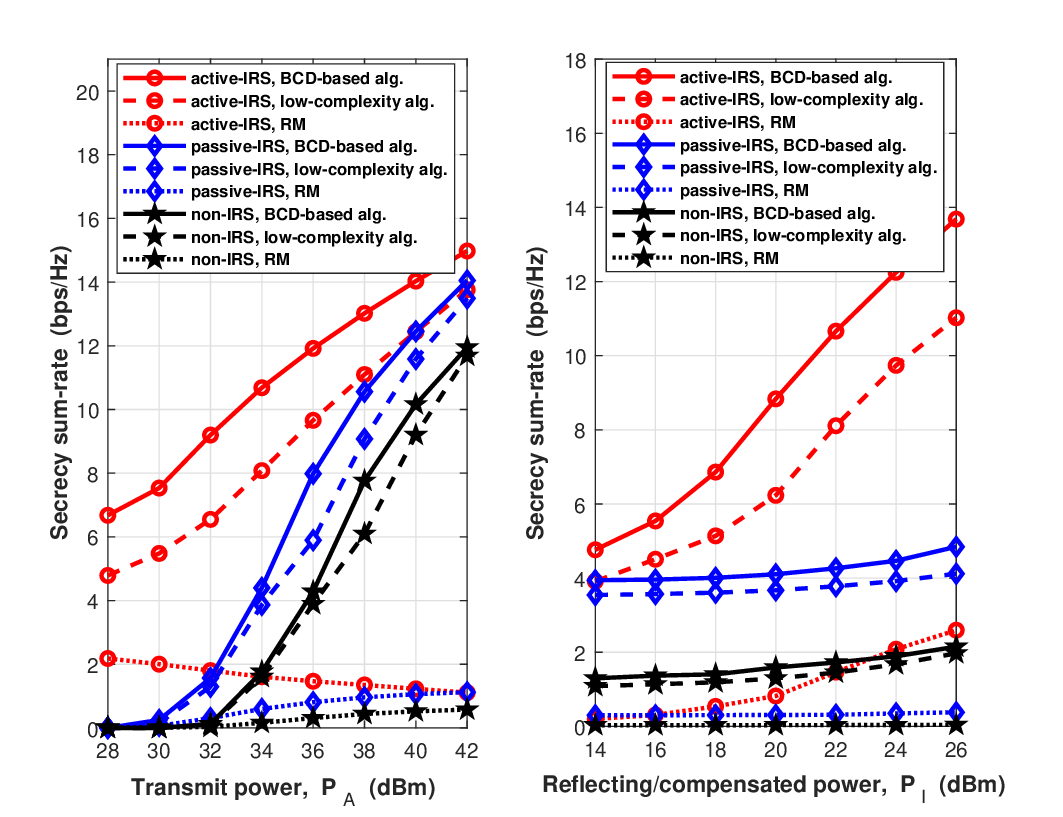}
	\caption{Secrecy sum-rate vs. (left) the Tx power and (right) the reflecting/compensated power (For the passive- and non-IRS schemes, $P_{\rm I}$ is compensated to the transmitter).}
	\label{versusTransP}
\end{figure}

The left panel of Fig.~\ref{versusTransP} shows the relationship between secrecy sum-rate and transmit power for different schemes and algorithms, with $P_{\rm I} = 22 \, \text{dBm}$ and $E_{\rm th} = -10 \,\text{dBm}$. We can see that the active IRS scheme outperforms the other schemes, especially at low transmit power levels. This observation is because the active-IRS can simultaneously adjust both the amplitude and phase of the incident signal, thereby offering greater configuration flexibility compared to the passive-IRS and non-IRS schemes. Moreover, when the transmit power is low, the active IRS can amplify the incident signals to compensate for the channel attenuation of the legitimate user, resulting in a higher secrecy sum rate. It is also noticeable that the performance gap between the active and passive IRS schemes becomes smaller as transmit power increases. This observation is intuitive, as the active IRS cannot compensate for signal attenuation too much at high transmit power due to the reflecting power constraint. In addition, the performance gap between the BCD-based and low-complexity algorithms diminishes at high transmit power levels. This occurs because, at high transmit power levels, preventing data leakage to eavesdroppers becomes critical for ensuring the PLS, and thus, the low-complexity solution, which leverages ZF precoding, performs more closely to the BCD-based solution. It is also observed that the performance of the RM method is inferior. This observation is expected because, without considering Eve, the system neither introduces AN nor constrains beamforming directions to counteract eavesdropping, leading to substantial information leakage. In addition, when the transmit power increases, the active-IRS RM scheme achieves a lower secrecy sum rate. This occurs because, without proper AN and beamforming design, the increase in transmit power strengthens the signal more for the eavesdropper than for the legitimate user, leading to a decrease in the secrecy sum rate.

The right panel of Fig.~\ref{versusTransP} illustrates the secrecy sum-rate versus the reflecting/compensated power for different schemes and algorithms, with $P_{\rm A} = 34 \, \text{dBm}$ and $E_{\rm th} = -10 \,\text{dBm}$. As stated before, the compensated power $P_{\rm I}$ is added at the transmitter for the passive-IRS and non-IRS schemes for fair comparisons. It can be seen that, as the reflecting power increases, the active IRS scheme significantly improves the secrecy sum rate, while the passive- and non-IRS schemes only yield a slight improvement. There are two main reasons for this observation. On the one hand, the active-IRS offers greater design flexibility compared to the passive- and non-IRS schemes due to its adjustable amplitude and phase, and the increase in reflecting power can further enhance this flexibility. On the other hand, the signal transmitted through the Alice-IRS-Bob link experiences path loss twice, whereas the signal transmitted through the IRS-Bob link only experiences single-hop path loss. Therefore, allocating extra power budget $P_{\rm I}$ to the IRS (the active IRS scheme) is more effective in enhancing the received signal power at the legitimate user, resulting in a more significant security performance gain. Additionally, it can be seen that the secrecy sum-rate performance of the RM method is poor since it does not consider Eve.

\begin{figure}[t]
	\centering
	\includegraphics[width=0.45\textwidth]{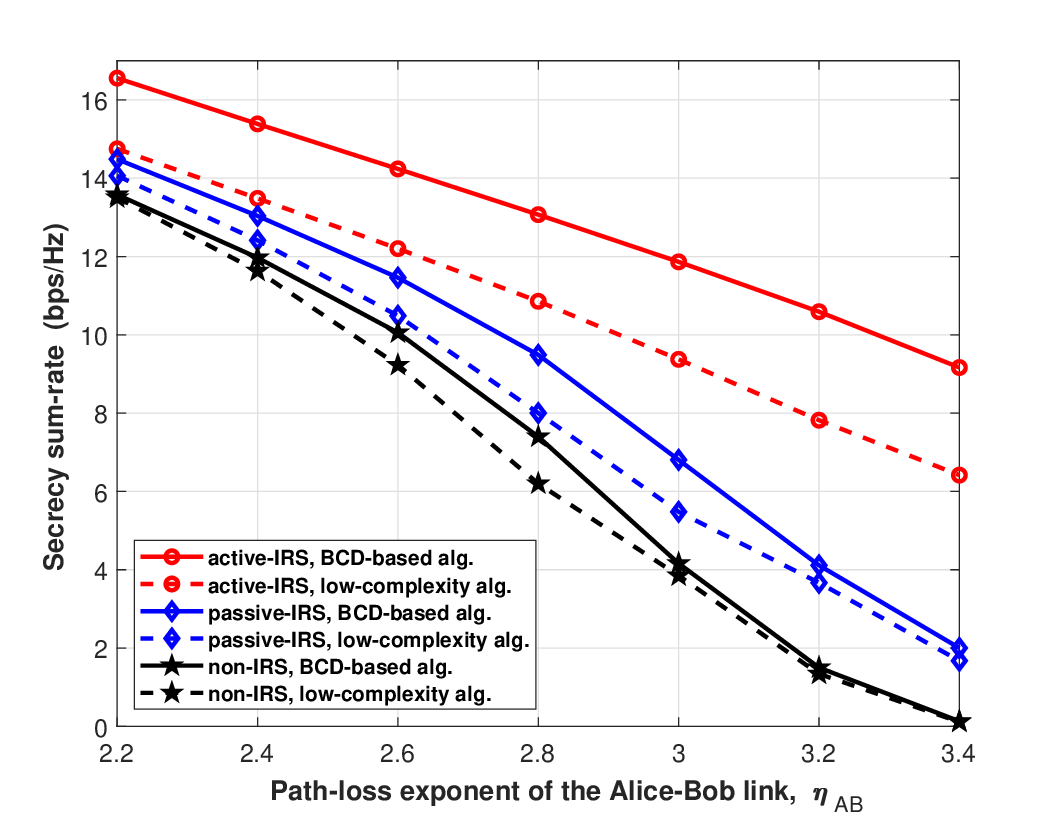}
	\caption{Secrecy sum-rate vs. the Alice-Bob link's path-loss exponent.}
	\label{versusExp}
\end{figure}

Fig.~\ref{versusExp} exhibits the secrecy sum-rate versus the path loss exponent of the link between Alice and Bob for different schemes and algorithms, with $P_{\rm A} = 34 \, \text{dBm}$, $P_{\rm I} = 22 \, \text{dBm}$, and $E_{\rm th} = -10 \,\text{dBm}$. As the channel quality between Alice and Bob worsens, the performance gap in security between the active and passive IRS becomes larger. This is intuitive because with a larger $\eta_{\rm AB}$, the direct link between Alice and Bob is more severely blocked, and as a result, the signal transmission relies more on the reflecting channel. Compared to the passive IRS scheme, the active IRS scheme can mitigate the path-loss attenuation of the reflecting channel, resulting in a slower decrease in the secrecy sum rate. 

\begin{figure}[t]
	\centering
	\includegraphics[width=0.45\textwidth]{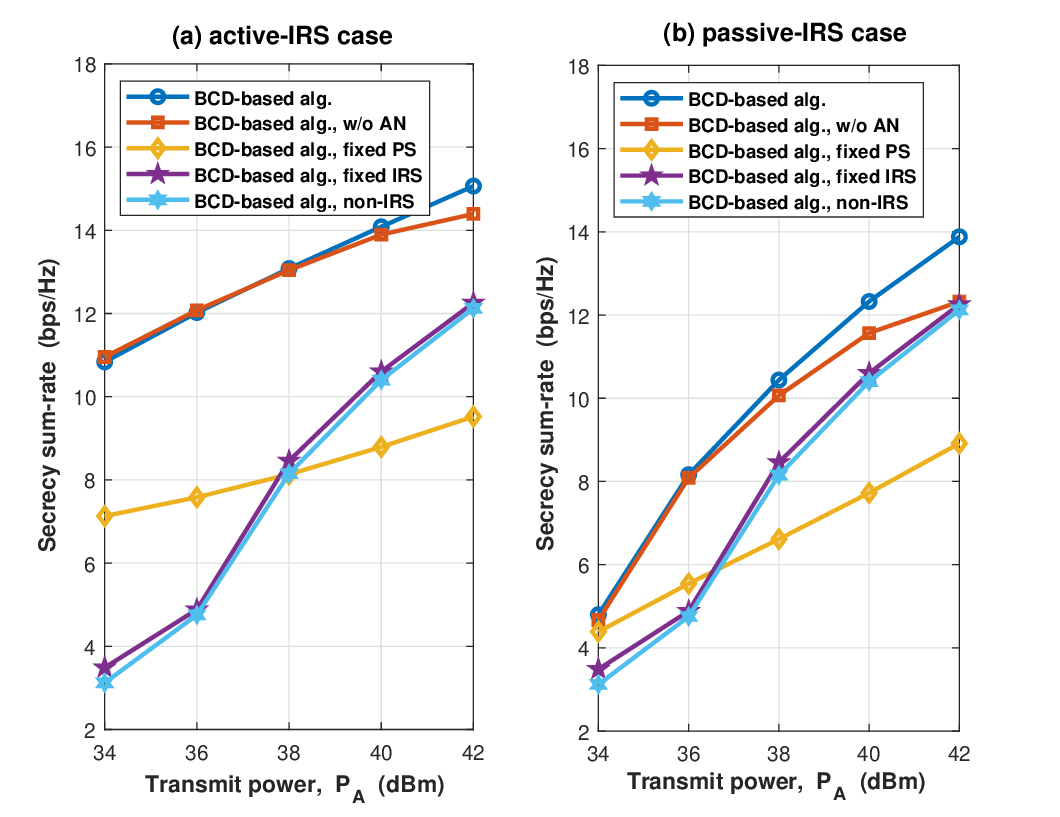}
	\caption{Secrecy sum-rate vs. the Tx power of Alice in the case of (left) active-IRS and (right) passive-IRS.}
	\label{optval}
\end{figure}

Finally, we explore the impact of optimization variables on the secrecy sum rate for the BCD-based algorithm, as shown in Fig.~\ref{optval}. In the legend, ``BCD-based alg., w/o AN" indicates that the AN precoding matrices are fixed as zero matrices, ``BCD-based alg., fixed PS" means that the PS factors are fixed at $0.1$, and ``BCD-based alg., fixed IRS" denotes that the IRS matrix is fixed as a random matrix. As the transmit power increases, the AN can significantly enhance performance for both active- and passive IRS schemes. It is because, with higher transmit power, optimizing the information precoding matrices alone is inefficient for restricting data leakage to the eavesdropping user, making incorporating AN precoding matrices beneficial. Furthermore, integrating the AN is more crucial for the passive IRS scheme than the active IRS scheme. It is because, for the active IRS scheme, the noise introduced at the reflecting array is controllable through the design of the IRS matrix and can thus act as a form of AN. In contrast, the thermal noise introduced at the IRS is negligible for the passive IRS scheme. In addition, we observe that the fixed-IRS scheme only brings a slight performance improvement compared to the non-IRS scheme, and the fixed-PS scheme performs worse when the transmit power becomes large.

\section{Conclusion}
\label{Sec-Concl}
The paper focuses on optimizing information precoding, power splitting, and IRS for an active-IRS-assisted MIMO-OFDM SWIPT system in the presence of an eavesdropper. The main objective is to maximize the secrecy sum rate while considering constraints such as maximum transmit power, maximum reflecting power, and minimum harvested energy. To address this complex non-convex problem, we first developed a BCD-based algorithm to find a sub-optimal solution, followed by a heuristic algorithm to achieve lower computational complexity. The simulation results demonstrate that the active IRS outperforms the passive and non-IRS systems in terms of secrecy sum rate, especially when the transmit power is low or the direct link is severely obstructed. Increasing the power budget at the active IRS can significantly improve the secrecy sum rate. Additionally, the active IRS scheme reduces the need to integrate AN at the transmitter, as the noise introduced at the active IRS can serve as AN. Despite the maximum secrecy sum rate, adopting secrecy energy efficiency as the objective and exploring the trade-off between the secrecy sum rate and total power consumption could be a future research topic.

\bibliographystyle{IEEEtran}

\bibliography{bibtex}

\begin{IEEEbiography}
	[{\includegraphics[width=1in,height=1.25in,clip,keepaspectratio]{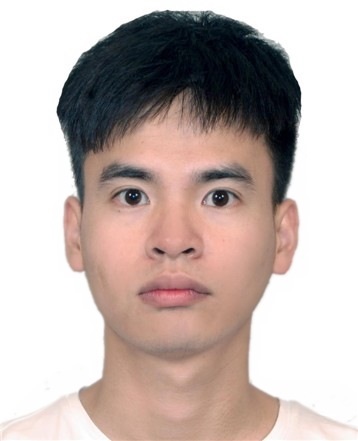}}]{Xingxiang Peng} received the B.Eng. degree from the School of Information Engineering, Guangdong University of Technology, Guangzhou, China, in 2018, and the M.Eng. degree from the School of Electronics and Information Technology, Sun Yat-sen University, Guangzhou, China, in 2020, where he is currently pursuing the Ph.D. degree in Electronics Science and Technology. His research interests include simultaneous wireless information and power transfer, intelligent reflecting surface-assisted communications, and optimization methods.
\end{IEEEbiography}

	\vspace{-10pt}

\begin{IEEEbiography}
	[{\includegraphics[width=1in,height=1.25in,clip,keepaspectratio]{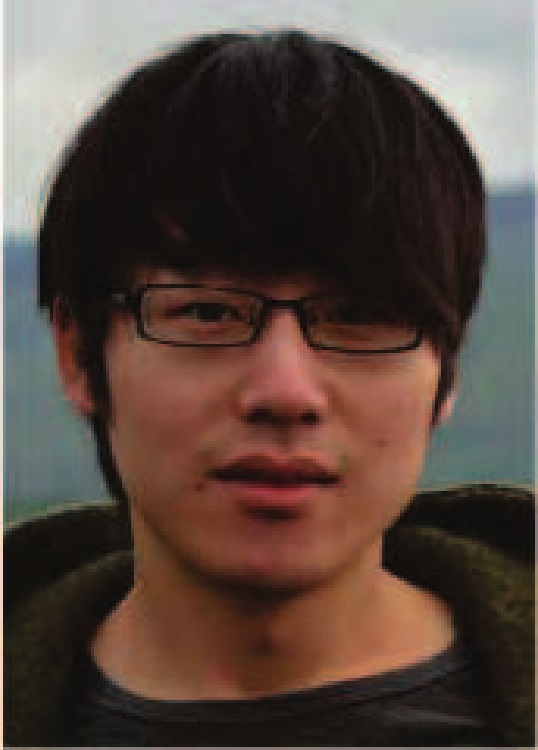}}]{Peiran Wu}
	(Member, IEEE) received a Ph.D. degree in electrical and computer engineering from The University of British Columbia (UBC), Vancouver, Canada, in 2015.
	
	From October 2015 to December 2016, he was a Post-Doctoral Fellow at UBC. In the Summer of 2014, he was a Visiting Scholar with the Institute for Digital Communications, Friedrich-Alexander-University Erlangen-Nuremberg (FAU), Erlangen, Germany. Since February 2017, he has been with Sun Yat-sen University, Guangzhou, China, where he is currently an Associate Professor. Since 2019, he has been an Adjunct Associate Professor with the Southern Marine Science and Engineering Guangdong Laboratory, Zhuhai, China. His research interests include mobile edge computing, wireless power transfer, and energy-efficient wireless communications. 
	
	Dr. Wu was a recipient of the Fourth-Year Fellowship in 2010, the C. L. Wang Memorial Fellowship in 2011, the Graduate Support Initiative (GSI) Award from UBC in 2014, the German Academic Exchange Service (DAAD) Scholarship in 2014, and the Chinese Government Award for Outstanding Self-Financed Students Abroad in 2014.
\end{IEEEbiography}

	\vspace{-10pt}
	
\begin{IEEEbiography}
	[{\includegraphics[width=1in,height=1.25in,clip,keepaspectratio]{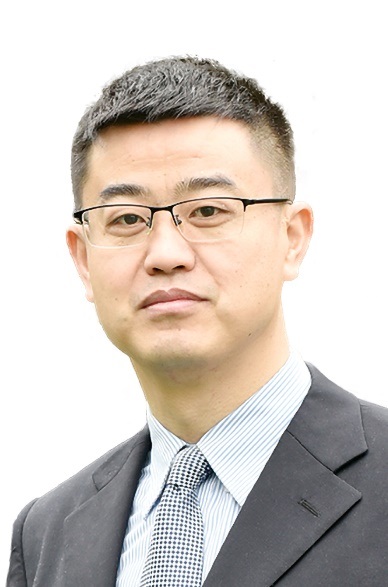}}]{Junhui Zhao} 
	(Senior Member, IEEE) received the M.S. and Ph.D. degrees from Southeast University, Nanjing, China, in 1998 and 2004, respectively. 
	
	From 1998 to 1999, he worked with Nanjing Institute of Engineers at ZTE Corporation. Then, he worked as an Assistant Professor in 2004 at the Faculty of Information Technology, Macao University of Science and Technology, and continued there till 2007 as an Associate Professor. In 2008, he joined Beijing Jiaotong University as an Associate Professor, where he is currently a Professor at the School of Electronics and Information Engineering. Meanwhile, he was also a short-term Visiting Scholar at Yonsei University, South Korea in 2004 and a Visiting Scholar at Nanyang Technological University, Singapore from 2013 to 2014. His current research interests include wireless communication, internet of things, and information processing in traffic.
\end{IEEEbiography}

	\vspace{-10pt}
	
\begin{IEEEbiography}
	[{\includegraphics[width=1in,height=1.25in,clip,keepaspectratio]{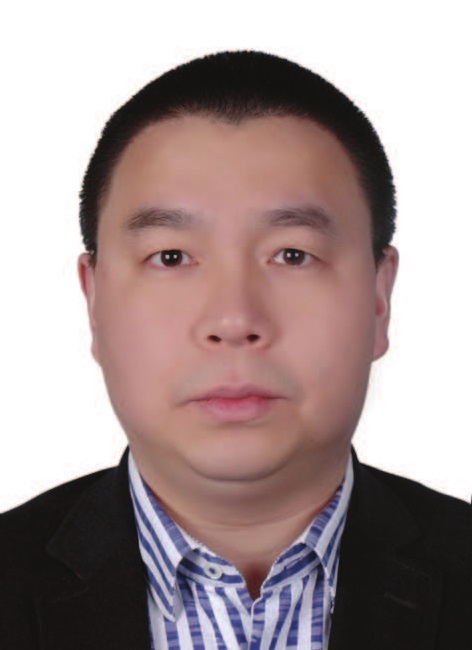}}]{Minghua Xia}
	(Senior Member, IEEE) received a Ph.D. degree in telecommunications and information systems from Sun Yat-sen University, Guangzhou, China, in 2007.
	
	From 2007 to 2009, he was with the Electronics and Telecommunications Research Institute (ETRI), South Korea, and with the Beijing Research and Development Center, Beijing, China, where he worked as a member and then as a Senior Member of Engineering Staff. From 2010 to 2014, he was in sequence with The University of Hong Kong, Hong Kong; King Abdullah University of Science and Technology, Jeddah, Saudi Arabia; and the Institut National de la Recherche Scientifique (INRS), University of Quebec, Montreal, Canada, as a Post-Doctoral Fellow. Since 2015, he has been a Professor at Sun Yat-sen University. Since 2019, he has also been an Adjunct Professor with the Southern Marine Science and Engineering Guangdong Laboratory, Zhuhai. His research interests are in the general areas of wireless communications and signal processing.
\end{IEEEbiography}

\vfill

\end{document}